\documentclass[mnsc, nonblindrev]{informs3} 

\OneAndAHalfSpacedXI

\usepackage{natbib}
\bibpunct[, ]{(}{)}{,}{a}{}{,}%
\def\bibfont{\small}%
\TheoremsNumberedThrough     
\ECRepeatTheorems

\EquationsNumberedThrough    

\usepackage{amsmath,amssymb,amsfonts}
\usepackage{mathtools}
\usepackage[noend]{algorithmic}
\usepackage{algorithm}
\usepackage{textcomp}
\usepackage{xcolor}	
\usepackage{graphicx}
\usepackage{booktabs}
\usepackage{multirow}
\usepackage{makecell}
\usepackage{tabularx}
\usepackage{latexsym}
\usepackage{subcaption}

\def\BibTeX{{\rm B\kern-.05em{\sc i\kern-.025em b}\kern-.08em
    T\kern-.1667em\lower.7ex\hbox{E}\kern-.125emX}}

\newlength\myindent
\usepackage[T1]{fontenc}
\usepackage{lmodern}
\usepackage[utf8]{inputenc} 
\usepackage[T1]{fontenc}    
\definecolor{DarkBlue}{rgb}{0,0.08,0.45}
\usepackage[backref = false, bookmarks, colorlinks = true, plainpages = false, citecolor = DarkBlue , urlcolor = DarkBlue, filecolor = DarkBlue, linkcolor = DarkBlue]{hyperref}    
\usepackage{url}            
\usepackage{booktabs}       
\usepackage{amsfonts}       
\usepackage{nicefrac}       
\usepackage{microtype}      
\usepackage{amsmath}
\usepackage{rotating}
\usepackage{mathrsfs}
\usepackage{array,multirow}
\usepackage[justification=centering]{caption}
\usepackage{bbm}

\usepackage{makecell}

\usepackage{tabularx}

\usepackage{mathtools}
\DeclarePairedDelimiterX{\norm}[1]{\lVert}{\rVert}{#1}

\usepackage{circuitikz}
\usepackage{tablefootnote}

\newtheorem{property}{Property}

\newcommand{\SWITCH}[1]{\STATE \textbf{switch} (#1)}
\newcommand{\ENDSWITCH}{\STATE \textbf{end switch}}
\newcommand{\CASE}[1]{\STATE \textbf{case} #1\textbf{:} \begin{ALC@g}}
\newcommand{\ENDCASE}{\end{ALC@g}}

\newcommand{\DEFAULT}{\STATE \textbf{default:} \begin{ALC@g}}
\newcommand{\ENDDEFAULT}{\end{ALC@g}}
\newcommand{\DEFAULTLINE}[1]{\STATE \textbf{default:} }

\def\tw#1{\begin{color}{black}{#1}\end{color}}

\def\minor#1{\begin{color}{black}{#1}\end{color}}

\begin{document}
\TITLE{\Large{Constant-Factor Algorithms \\for Revenue Management with Consecutive Stays}}
\ARTICLEAUTHORS{
 \AUTHOR{Ming Hu}
 \AFF{Rotman School of Management, University of Toronto, \EMAIL{ming.hu@rotman.utoronto.ca}}
\AUTHOR{Tongwen Wu}
\AFF{Rotman School of Management, University of Toronto, \EMAIL{tw.wu@rotman.utoronto.ca}} 
 }

\ABSTRACT{ 
We study network revenue management problems motivated by applications such as railway ticket sales and hotel room bookings. Requests, each requiring a resource for a consecutive stay, arrive sequentially with known arrival probabilities. We investigate two scenarios: the accept-or-reject scenario, where a request can be fulfilled by assigning any available resource; and the BAM-based scenario, which generalizes the former by incorporating customer preferences through the basic attraction model (BAM), allowing the platform to offer an assortment of available resources from which the customer may choose. \tw{We develop polynomial-time policies and evaluate their performance using approximation ratios, defined as the ratio between the expected revenue of our policy and that of the optimal online algorithm. When each arrival has a fixed request type (e.g., the stay interval is fixed), we establish constant-factor guarantees: a ratio of $1 - 1/e \approx 0.632$ for the accept-or-reject scenario and \minor{$0.271$} for the BAM-based scenario. We further extend these results to the case where the request type is random (e.g., the stay interval is random). In this setting, the approximation ratios incur an additional multiplicative factor of $1 - 1/e$, resulting in guarantees of at least $0.399$ for the accept-or-reject scenario and \minor{$0.171$} for the BAM-based scenario. These constant-factor guarantees stand in sharp contrast to the prior nonconstant competitive ratios that are benchmarked against the offline optimum.}


}

\maketitle 

\vspace{-2em}
\section{Introduction}
The network revenue management (NRM) problem is a classic resource allocation challenge in which a decision-maker must manage limited resources while deciding over time whether and how to fulfill sequentially arriving requests, each of which may require multiple resources. NRM has been widely applied in various settings, including the travel and hospitality industries \citep{gallego2019revenue}.
\tw{In this work, we study NRM problems where each resource has an ordered sequence of slots, and each request requires a contiguous interval of slots from a single resource.}
To motivate our problem, we present two illustrative scenarios.

\begin{figure}
\centering
\caption{Illustrative Examples}
\begin{subfigure}{.5\textwidth}
  \centering
  \includegraphics[width=.7\linewidth]{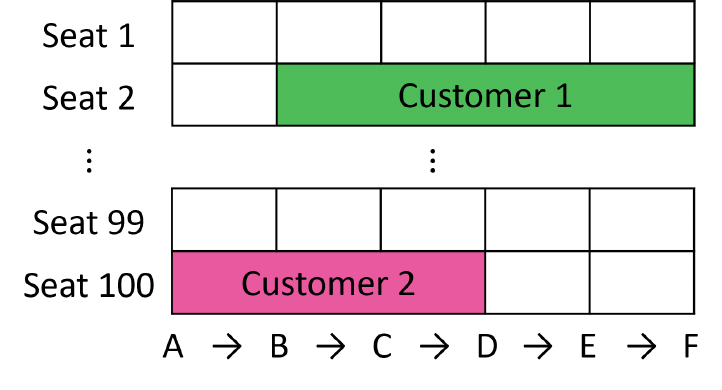}
  \caption{Seat Availability Status}
  \label{fig:seat}
\end{subfigure}%
\begin{subfigure}{.5\textwidth}
  \centering
  \includegraphics[width=.7\linewidth]{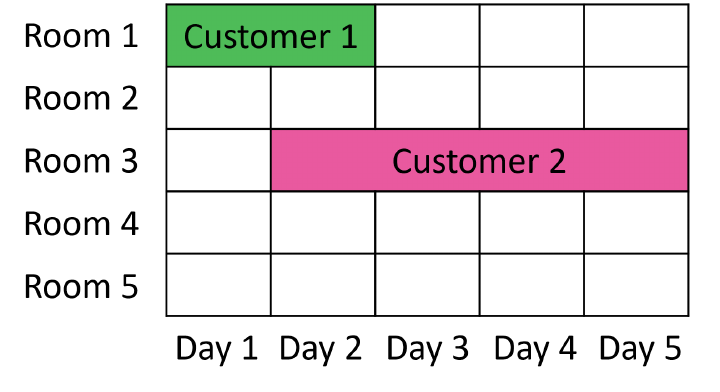}
  \caption{Hotel Room Reservation Status}
  \label{fig:room}
\end{subfigure}
{\footnotesize \textit{Note.} For both subfigures, blank slots indicate availability, whereas colorful slots represent reservations made by two distinct customers, respectively.}
\label{fig:example}
\end{figure}

First, in the high-speed railway industry, which has experienced rapid global expansion \citep{globenewswire2025railroads}, seat assignment plays a critical role in maximizing profits \citep{zhu2023assign}.
Consider, for example, a train route running from Station A to Station F, with intermediate stops at Stations B, C, D, and E. The train has 100 numbered seats, and each passenger requests a seat for either the full route or a segment of it (see Figure~\ref{fig:seat}). The resource allocation in this context can be visualized as a Gantt chart, where each row represents a seat and each column corresponds to a leg of the journey. For every online booking request, the decision-maker must determine not only whether to accept the request but also which seat to assign, ensuring that no seat is allocated to overlapping itineraries on the same leg.


Second, the scenario described above can be extended to revenue management under customer choice \citep{rusmevichientong2023revenue}. This applies to platforms such as Airbnb and boutique hotels, where resources such as rooms or houses are uniquely defined by features such as decor, views, and prices (see Figure~\ref{fig:room}). Customers arrive with booking requests for specific future time intervals, and resources are available only during the remaining time slots after blocking those reserved by previous customers. A key operational challenge in this setting is that the decision-maker must present an assortment of available resources for the customer to select from, based on a specified choice model. Effectively managing these bookings requires offering personalized assortments that strategically balance immediate revenue potential with future resource availability.


\subsubsection*{Our Model.}
In this paper, we study a class of network revenue management problems involving $M$ resources, each with $N$ initially available slots, where the slots may represent time-based capacities such as legs or days. The system operates over a finite planning horizon of $T$ periods, during which at most one customer request may arrive per period. 
\tw{Throughout the paper, the arrival event and the request type in each period are independent across periods and are drawn from known distributions.
A request may have a known type, specified by its arrival probability, the required stay interval, and the rewards for assigning it to each resource, which we refer to as \textit{Bernoulli arrivals}. More broadly, the request type may be random, with both the required interval and the associated rewards drawn from known distributions, which we refer to as \textit{general arrivals}.}
We consider two scenarios that differ in how allocation decisions are made: the \textit{accept-or-reject scenario} and the \textit{BAM-based scenario}. In the accept-or-reject scenario, akin to seat assignment in railway ticketing, the decision-maker must not only decide whether to accept a request but also assign an appropriate interval of available slots. In the BAM-based scenario, the problem becomes more complex because a customer's choice is modeled using the basic attraction model \citep{luce1959individual}.
In this case, the decision-maker must offer an assortment of resources at given prices, from which a customer probabilistically selects a resource, making the allocation decision more intricate than in the accept-or-reject setting.
Indeed, the BAM-based scenario is a generalization of the accept-or-reject scenario. 
The objective in both scenarios is to design a policy that maximizes the total expected revenue over the planning horizon.


\subsubsection*{Theoretical Results.}
\tw{We extend ideas from the online stochastic matching to the network revenue management setting and obtain constant-factor approximation ratios for the first time. 
In the NRM literature, there has been a recent effort to push the limits of performance guarantees.
The accept-or-reject scenario under general arrivals was studied by \cite{zhu2023assign}. 
They compared the performance of their policy to an \textit{offline optimum}, that is, a benchmark with full knowledge of the realized request arrivals. 
However, their guarantees are asymptotic and rely on stationarity of the arrival distribution.
Indeed, the static bundle pricing mechanism proposed by \cite{chawla2019pricing} already implies a competitive ratio\footnote{\tw{Throughout the paper, we differentiate the terms competitive ratio and approximation ratio. 
The approximation ratio $\alpha$ means that a polynomial-time policy achieves at least an $\alpha$ fraction of the value of the online optimum. 
The competitive ratio $\alpha$ means that a possibly non-polynomial-time policy achieves at least an $\alpha$ fraction of the value of the offline optimum.}} of $O(\log \log L/\log L)$ for this setting, where $L$ is the maximum length of the requested intervals.
For a more general setting under general arrivals, where customer choice follows a general choice model, including the BAM model, 
\cite{rusmevichientong2023revenue} propose a polynomial-time policy and benchmark it against the \textit{online optimum}, a benchmark that knows the distribution of future arrivals but not their realizations, and is therefore more approachable than the offline optimum. 
Their analysis yields an approximation ratio of only $\Omega(1/L)$.
Later, \cite{simchi2025greedy} make significant progress by proving a competitive ratio of $\Omega(1/\log L)$ when even compared to the offline optimum for this general setting.\footnote{\tw{We highlight two points regarding the result in \cite{simchi2025greedy}. 
First, their guarantee is derived for general choice model-based scenarios, which include the BAM-based scenario, and their offline optimum benchmark assumes that the decision-maker knows the realized request type but not the realization of the random choice. 
Second, their analysis is carried out under adversarial arrivals, meaning that even the distribution of future arrivals is unknown. As a result, their competitive ratio necessarily includes a nonconstant factor that depends on the magnitude of the rewards, in addition to the term $1 / \log L$. Nevertheless, their bucketing idea, which groups intervals by length, plays a key role in their proof and can be adapted to the setting with known arrival distributions to remove this nonconstant dependence. Therefore, in Table~\ref{tbl:results}, we specify their guarantee for the BAM-based scenario under general arrivals, though their results also apply to the general choice model-based scenario and to adversarial arrivals. 
}} 
On the other hand, \cite{im2011secretary} show that even with a single resource ($M = 1$), the accept-or-reject scenario under general arrivals cannot achieve a competitive ratio better than $O(\log \log L / \log L)$. Existing results are listed in Table~\ref{tbl:results}.
Hence, a key open question remains: can we obtain constant-factor approximation ratios independent of $L$ when measured against the \textit{online optimum}?
}

\begin{table}[!htbp]
\tw{
\caption{Comparison of Results under Different Arrivals, Scenarios, and Benchmarks}
\label{tbl:results}
\small
\centering
\begin{tabular}{c c c c c}
\toprule
& \multicolumn{2}{c}{Accept-or-reject scenario} 
& \multicolumn{2}{c}{BAM-based scenario} \\
\cmidrule(lr){2-3} \cmidrule(lr){4-5}
Benchmark 
& Bernoulli arrivals 
& General arrivals 
& Bernoulli arrivals 
& General arrivals \\
\midrule
 \multirow{2}{*}{Offline optimum}& - & $\le O({\log\log L}/{\log L})$  &- & $\le O({\log\log L}/{\log L})$ \\
& \multicolumn{2}{c}{$\ge \Omega({\log \log L}/{\log L})$ \citep{chawla2019pricing}} & \multicolumn{2}{c}{$\ge\Omega(1/\log L)$  \citep{simchi2025greedy}}
 \\
 \midrule 
 \multirow{2}{*}{Online optimum} & \multicolumn{4}{c}{ $\le 0.95+\epsilon$ (Proposition~\ref{prop:hardness_accept_reject}) } \\
& $\ge 1 - 1/e$ (Thm.~\ref{thm:1 - 1/e}) 
& $\ge (1 - 1/e)^2$ (Thm.~\ref{thm:accept_or_reject_general}) 
& \minor{$\ge 0.271$} (Thm.~\ref{thm:1/8}) 
& \minor{$\ge 0.271(1 - 1/e)$} (Thm.~\ref{thm:BAM_general}) \\
\bottomrule
\vspace{0.1cm}
\end{tabular}
{\footnotesize \textit{Note.} Both impossibility results when benchmarking policies against the offline optimum follow from Theorem~3.1 in \cite{im2011secretary}. For the offline version of the accept-or-reject scenario, the best-known approximation ratio is $1 - 1/e$ attributed to \cite{bhatia2007algorithmic}. 
Beyond the BAM model, the best known competitive ratio for the general choice model-based scenario remains $\Omega(1/\log L)$, as shown by \cite{simchi2025greedy}.
}
}
\end{table}

\tw{The online optimum is less studied because it is defined through a high-dimensional dynamic program.
Its study was recently initiated by \cite{papadimitriou2024online} in the context of online stochastic matching, which can be viewed as the $L=1$ case of the accept-or-reject scenario with general arrivals. 
They broke the long-standing $0.5$ barrier for the best competitive ratio by establishing a $0.51$-approximation ratio. 
This ratio was later sequentially improved to $0.678$ by \cite{saberi2021greedy, braverman2022max, naor2025online, braverman2025new}. 
Notably, \cite{papadimitriou2024online} and \cite{braverman2025new} further show that there exists a constant less than one (but greater than $0.99$) such that approximating the online optimum within this constant is PSPACE-hard.}

We extend the \textit{proposal-discarding algorithm} developed by \cite{braverman2025new} to our NRM problems, and derive constant-factor approximation ratios when comparing with the online optimum. 
These results are summarized in Table~\ref{tbl:results}.
First, for the accept-or-reject scenario under Bernoulli arrivals, an extension of the proposal-discarding algorithm (Algorithm~\ref{alg:single_item}) that exploits the decomposable property in the single-resource case yields a $(1 - 1/e)$-approximation ratio. 
\tw{This already matches the best-known $(1 - 1/e)$-approximation ratio for the offline version of this problem \citep{bhatia2007algorithmic}, where the goal is to find an allocation that maximizes total revenue given the realized requests. 
Moreover, we complement this result by showing that the offline version is NP-hard to approximate within a factor of $0.95 + \epsilon$ for any $\epsilon > 0$. 
This hardness immediately applies to all online settings, providing an upper bound on approximability that is farther from one than the PSPACE-hardness results of \cite{papadimitriou2024online} and \cite{braverman2025new}. 
}
Second, we extend our analysis to settings with customer choice and general arrivals. 
For the BAM-based scenario under Bernoulli arrivals, we generalize Algorithm~\ref{alg:single_item} by using a sales-based fluid relaxation of the online optimum and developing a new coupling procedure, which together yield a \minor{$0.271$}-approximation ratio through Algorithm~\ref{alg:assortment_based}. 
\tw{To handle general arrivals, we further generalize Algorithm~\ref{alg:assortment_based} by incorporating an attenuation mechanism, and show that the resulting algorithm achieves constant-factor guarantees with an additional multiplicative factor of $(1 - 1/e)$ for both scenarios.
}

\subsubsection*{Technical Contributions.}
Our key technical contributions, along with the specific challenges they address, are summarized as follows.


\paragraph{Decomposable property.}
A central challenge in network revenue management is the difficulty introduced by dependencies on the parameter $L$, which represents the maximum length of requested intervals. This difficulty stems from the need to simultaneously track the availability of up to $L$ slots, particularly when resource requirements have arbitrary structures, as prior work has revealed \citep{ma2024online}. In contrast, our setting assumes that requested slots are consecutive, enabling us to exploit this structural feature for a more compact representation of resource availability.
Specifically, we represent the state of each resource using the notion of maximal sequences of available slots. Importantly, the number of such sequences is bounded by $O(N^2)$, where $N$ is the initial number of slots per resource. Although a resource's availability status may initially involve multiple maximal sequences, we uncover an underlying decomposable property that substantially simplifies the design of approximation algorithms.
To illustrate this insight, we begin with a warm-up case in Section~\ref{sec:warm_up}, focusing on a single-resource accept-or-reject scenario. We demonstrate that this simpler setting admits an optimal polynomial-time policy via dynamic programming, where states are defined in terms of maximal sequences. Building on this foundation, we introduce variables corresponding to maximal sequences in our fluid relaxation framework for the general case.


\paragraph{Proposal-discarding algorithm.}
We evaluate our policies by benchmarking them against the online optimum, which can be computed via dynamic programming. However, this approach typically incurs exponential state complexity when multiple resources are involved. 
Our algorithm for the accept-or-reject scenario draws inspiration from the proposal-discarding framework introduced by \cite{braverman2022max,braverman2025new} in the context of online stochastic matching. 
\tw{In their setting, an optimal solution to a fluid relaxation of the online optimum is first computed.
\cite{braverman2022max} propose that each unmatched offline node independently submits a proposal to match with the arriving online node, using probabilities derived from the fluid solution, and the max-weight proposal is then selected for matching.
Later, \cite{braverman2025new} extend this by adding a discarding step, in which the remaining proposals are matched to virtual copies of the online node, each with an independent arrival.
This procedure ensures probabilistic independence across resources.
To adapt this framework to our setting, we generalize from the binary availability status of nodes to a richer representation based on the maximal sequences of each resource. 
Specifically, we introduce the notion of virtual resource status, which may be marked as unavailable even when the actual resource is available (reflecting discarding). 
This design allows each resource to submit proposals independently based on its local availability state, thereby preserving probabilistic independence across resources and enabling scalable algorithmic implementation.
}


\paragraph{Policy design for the BAM-based scenario.}
Extending the proposal-discarding algorithm to the BAM-based scenario introduces a unique challenge: the decision-maker can influence customer behavior only indirectly through the assortment of resources offered. To the best of our knowledge, this work is the first to integrate choice models into the proposal-discarding algorithmic framework.
Our approach incorporates three new techniques. 
First, drawing inspiration from the sales-based linear program (SBLP) introduced by \citet{gallego2015general}, we formulate a polynomial-sized fluid relaxation that approximates the online optimum. 
Second, since the decision here is an offered assortment, we construct the assortment by selecting the subset of the proposal set that maximizes the expected revenue. This can be viewed as a natural generalization of selecting the max-weight proposal in the original proposal-discarding algorithm. 
Third, we observe that, unlike in the accept-or-reject scenario, probabilistic independence across resources does not hold in general, because a customer's choice depends on the entire assortment. 
\tw{To address this, we introduce a technical mechanism: a randomized coupling function that links the customer's choice to a randomly generated subset of resources, designed to restore probabilistic independence. 
In particular, given two vectors $\mathbf{q}, \mathbf{q}' \in [0,1]^{M}$, we couple a random choice $\tilde{j}$ satisfying $\Pr[\tilde{j} = j] = q'_j$ with a random set $\mathcal{Q}$ that includes each $j \in [M]$ independently with probability $q_j$, in such a way that $\tilde{j} \in \mathcal{Q}$ whenever $\tilde{j} \in [M]$.\footnote{We use $[n]$ to denote $\{1,\dots, n \}$ for any nonnegative integer $n$.}
}

\tw{We develop new performance analysis for the BAM-based scenario. In short, we lower-bound the expected revenue of the highest-revenue assortment constructed from the proposal set, in which each resource is included independently. We partition resources into small- and large-attractiveness groups and derive tighter lower bounds on the expected revenue of the highest-revenue assortment in each group. We then combine these two bounds to obtain the final lower bound.
}

\tw{
\paragraph{Extension to general arrivals.}
Although the proposal-discarding algorithm for online matching can be extended from Bernoulli arrivals to general arrivals without any loss in the approximation ratio \citep{braverman2022max, braverman2025new}, this extension relies critically on the matched/unmatched status of each node, which allows the analysis to move beyond probabilistic independence and exploit certain forms of negative dependence. 
In our setting, the resource status is much more complex, and it is not even clear what kind of negative dependence, if any, should hold. 
To handle general arrivals, we generalize the algorithm for Bernoulli arrivals by introducing attenuation factors, each specifying that an arriving customer of a given type is offered the assortment only with a certain probability. 
This construction allows us to couple the customer's random choice to the virtual resource's status via a two-dimensional generalization of the randomized coupling function developed earlier. We show that introducing such an attenuation mechanism incurs an additional multiplicative factor of $1 - 1/e$ in the approximation ratios.}

\subsubsection*{Organization.} 
\tw{In Section~\ref{sec:literature}, we provide a further review of related work. Section~\ref{sec:model} introduces our model, covering both the accept-or-reject and BAM-based scenarios under Bernoulli arrivals, and Section~\ref{sec:warm_up} includes a warm-up analysis of the optimal policy for the single-resource case. 
We present our results for these two scenarios under Bernoulli arrivals in Sections~\ref{sec:single_item} and \ref{sec:assortment}, respectively. 
In Section~\ref{sec:general_arrivals}, we extend our results to general arrivals.}


\section{Literature Review} 
\label{sec:literature}
We review three streams of literature to position our work.

\subsubsection*{Network Revenue Management.}
The network revenue management (NRM) problem is named after the underlying network structure of resources, in which fulfilling a single request typically involves allocating multiple resources simultaneously \citep{gallego1997multiproduct}. 
While demand distributions are typically assumed to be known in advance, computing an optimal policy via dynamic programming is often computationally intractable due to the curse of dimensionality.
\tw{A considerable body of research has focused on developing approximate dynamic programming methods to address this challenge. Starting from \cite{talluri1998analysis}, which uses the bid price method to approximate the value function, much of the subsequent literature has aimed to construct increasingly accurate value function approximations (see, e.g., \citealt{bertsimas2003revenue}, \citealt{adelman2007dynamic}, and \citealt{topaloglu2009using} for the accept-or-reject scenario, as well as \citealt{zhang2009approximate} and \citealt{kunnumkal2010new} for the choice-based scenario).  
Another major line of work establishes the asymptotic optimality of simple policies that periodically re-solve a deterministic linear program obtained by replacing random variables with their expectations (see, e.g., \citealt{reiman2008asymptotically, jasin2012re, jasin2013analysis, bumpensanti2020re}, and the recent survey by \citealt{balseiro2024survey}).
}

More recently, worst-case analysis has received increasing attention, with studies evaluating policies relative to either the optimal dynamic solution \citep{ma2020approximation, rusmevichientong2023revenue} or the offline optimum with full knowledge of future demand \citep{baek2022bifurcating, ma2024online}. These papers typically establish performance guarantees that scale as $O(1/L)$, where $L$ denotes the maximum number of resources required per request.
A notable exception is \cite{simchi2025greedy}, which achieves improved performance bounds under a specific structure of requested resources. In contrast, our work is the first to establish constant-factor approximation ratios under this structure, thereby advancing the worst-case analysis of NRM.

\tw{
\subsubsection*{Interval Scheduling.}
The interval scheduling problem is another line of work related to ours. We refer readers to \cite{kolen2007interval} and \cite{kovalyov2007fixed} for comprehensive reviews of interval scheduling and its variants. Below, we highlight the results most closely connected to our setting, divided into offline and online categories.
}

\tw{The offline interval scheduling problem was introduced by \cite{arkin1987scheduling}. In this problem, one must schedule jobs (requests) on machines (resources), where each job has a fixed start time, end time, and value. A machine can process at most one job at a time, and the goal is to select a feasible schedule that maximizes total value. They show that when machines are identical, the problem is solvable in polynomial time, while the variant in which each job can be scheduled only on a subset of machines is NP-hard, with a reduction from 3-SAT (see Definition~\ref{dfn:3-sat}). Both cases are special instances of the offline version of the accept-or-reject scenario, and our stronger hardness result (Proposition~\ref{prop:hardness_accept_reject}) applies to the latter case. In fact, our proof follows a similar reduction but adjusts the weights and is a gap-preserving reduction. The best known approximation ratio for this setting is $1 - 1/e$, given by \cite{bhatia2007algorithmic}. Their algorithm can also be extended to the case where a job's value depends on the machine on which it is scheduled, which is exactly the offline version of the accept-or-reject scenario. We attribute the offline result to them, even though we do not explicitly describe this extension, since our policy already implies it.
}

\tw{The online interval scheduling problem was first studied by \cite{lipton1994online} and later examined under various settings. The closest work to ours is \cite{chawla2019pricing}, which studies an online Bayesian model with a single machine, where each job can choose from a set of possible intervals, each associated with a reward. Our accept-or-reject scenario under general arrivals is a special case of their model (by a reduction that concatenates our resources into a single resource), so their $O(\log \log L / \log L)$ competitive ratio applies to our setting. 
In addition, \cite{im2011secretary} give a tight upper bound of $O(\log \log L / \log L)$ on the competitive ratio for this model even when each job has only a single interval, and this upper bound also applies to our setting, since this single-interval model is a special case of our accept-or-reject scenario under general arrivals.
}

\subsubsection*{Online Bipartite Matching.}
Online bipartite matching has long been a foundational problem in the study of online algorithms. In the context of adversarial arrivals, one of the most influential results is due to \cite{karp1990optimal}, who established a seminal competitive ratio of $1 - 1/e$. For an overview of the extensive literature on related variants, we refer readers to the survey by \cite{mehta2013online}.
In settings where arrivals follow known distributions, a prominent research direction focuses on \textit{prophet inequalities}, which benchmark online algorithms against the offline optimum. This line of work has gained considerable attention in Bayesian selection and matching problems; see the comprehensive review by \citet{ma2024randomized}. 

\tw{More recently, there has been growing interest in benchmarking policies against the online optimum (often referred to as the \textit{philosopher inequality}). See \cite{huang2024online} for a review that simultaneously covers adversarial arrivals, prophet inequalities, and philosopher inequalities.
 This direction on the matching problems was initiated by the breakthrough of \cite{papadimitriou2024online}, with subsequent advances by \cite{saberi2021greedy}, \cite{braverman2022max}, \cite{naor2025online}, and \cite{braverman2025new}. 
The currently prevalent approach is the proposal-discarding algorithm framework introduced by \cite{braverman2025new}, which employs pivotal sampling to generate proposals and shows that it can achieve better approximation ratios than independent proposals. 
Our method extends this framework to accommodate more complex resource-availability structures, and, due to this added complexity, we continue to use independent proposals. 
In addition, we incorporate customer choice models, thereby broadening the applicability of the proposal-discarding approach to assortment optimization settings.
Parallel developments in comparing policies against the optimal online benchmark
have also appeared in other related problems, including variants of the matching problem (see, e.g., \citealt{anari2019nearly, feng2021two, dutting2023prophet, braun2024approximating, pollner2025optimal, sun2025combinatorial}), matching in continuous-time models (see, e.g., \citealt{aouad2020dynamic, kessel2022stationary, patel2024combinatorial, amanihamedani2024improved, amanihamedani2025adaptive}), and matching under an unknown arrival order \citep{sun2025online}.
}

\tw{One notable extension of online bipartite matching is online assortment optimization, where in each period the decision-maker offers an assortment of offline nodes to an online node, which then makes a choice according to a choice model. 
\cite{golrezaei2014real}, \cite{ma2020algorithms}, and \cite{goyal2025asymptotically} study this problem under adversarial arrivals. 
\cite{ma2021dynamic} and  \cite{feng2022near} consider a Bayesian setting and apply techniques from prophet inequalities. 
This Bayesian setting includes our BAM-based scenario under general arrivals with $L = 1$, and \cite{ma2021dynamic} obtain a $1/2$-competitive ratio in that case. 
However, since we compare policies against the online optimum and use a proposal-discarding framework for the philosopher inequality, our techniques for the BAM-based scenario differ from those used for adversarial arrivals and prophet inequalities.
}

\section{Model}
\label{sec:model}

We consider a generic resource allocation problem involving $M$ heterogeneous resources, denoted by $\mathcal{M} \triangleq \{1, \dots, M\}$. For any integers $i$ and $j$, we use the notation $[i, j]$ to represent the set $\{i, i+1, \dots, j\}$ if $i \le j$, and $\emptyset$ otherwise. Each resource is initially available during the set of time slots $\mathcal{N} \triangleq [1, N]$.
This notation accommodates a variety of applications. For example, in the context of railway ticket sales, the resources correspond to seats, and the time slots correspond to legs between consecutive stations. In hotel room bookings, the resources represent rooms or houses, and the time slots correspond to days.
We consider a finite planning horizon consisting of $T$ discrete time periods, where each period is sufficiently short to ensure at most one request arrives. We analyze two distinct scenarios, differentiated by the structure of the requests, as described below.


\subsection{Accept-or-Reject Scenario}
We begin with a scenario in which the decision-maker must respond to each incoming request by either rejecting it or accepting it along with an allocation of available time slots.
Formally, under \textit{Bernoulli arrivals}, in each period $t \in \mathcal{T} \triangleq \{1, \dots, T\}$, the request is characterized by a type $\theta_t = (p_t, l_t, r_t, \{w_{tj} \mid j \in \mathcal{M} \})$, where the request arrives independently with probability $p_t$, it demands consecutive time slots $[l_t, r_t] \subseteq \mathcal{N}$, and for each resource $j \in \mathcal{M}$, assigning the interval $[l_t, r_t]$ of resource $j$ yields a reward of $w_{tj} \ge 0$.
If the request arrives in period $t$, the decision-maker may choose to assign the requested time interval $[l_t, r_t]$ on any available resource $j \in \mathcal{M}$, thereby earning revenue $w_{tj}$. Alternatively, the request may be rejected, in which case no revenue or cost is accrued. 
One example for this scenario from \cite{zhu2023assign} is assigning seats to customers for selling train tickets: the resources \(\mathcal{M}\) correspond to seats, and each time slot represents a travel leg from one station to the next. In this context, \(w_{tj}\) denotes the ticket price from the departure station of leg \(l_t\) to the arrival station of leg \(r_t\).
In Section~\ref{sec:general_arrivals}, we extend the Bernoulli arrivals to \textit{general arrivals}, where $l_t,r_t,$ and $\{ w_{tj} |j \in \mathcal{M}\}$ can also be random and follow known distributions.


An instance of this scenario is denoted by $\mathcal{I} \triangleq \{\mathcal{M}, \mathcal{N}, \{ \theta_t \}_{t \in \mathcal{T}} \}$, and the problem for this instance can be formulated as follows:
\begin{align}
\label{eq:problem}
\overline{V}(\mathcal{I}) \triangleq \max_{\pi  } V^{\pi}(\mathcal{I}),
\end{align}
where $V^{\pi}(\mathcal{I})$ is the total expected revenue over the planning horizon under a feasible policy $\pi$, and $\overline{V}(\mathcal{I})$ represents the maximum achievable expected revenue across all feasible policies.
A policy is deemed feasible if its decisions rely solely on prior information and the observed history up to the current period. Our objective is to design a polynomial-time policy that approximately maximizes the expected total revenue.
We evaluate the quality of a policy by comparing it to the optimal policy, as formalized below:
\begin{definition}[Approximation Ratio]
A policy $\pi$ is said to be an $\alpha$-approximation if, for any instance $\mathcal{I}$, it runs in polynomial time and satisfies
$     V^\pi(\mathcal{I}) \ge \alpha  \cdot \overline{V}(\mathcal{I}).
    $
\end{definition}


\subsection{BAM-Based Scenario}

We further generalize the previous scenario by considering settings in which the decision-maker responds to each customer's arrival by offering an assortment of available resources, where the customer's selection is governed by a discrete choice model.
We assume that customer behavior follows a basic attraction model (BAM) \citep{luce1959individual}. Specifically, in each period $t \in \mathcal{T}$, under \textit{Bernoulli arrivals}, the request is characterized by a type $\theta_t = (p_t, l_t, r_t, \{w_{tj} \mid j \in \mathcal{M}\}, \{v_{tj} \mid j \in \mathcal{M}^+\})$, where $\mathcal{M}^+ \triangleq \mathcal{M} \cup \{0\}$ and index 0 represents the outside option. 
The parameters $p_t,l_t,r_t,\{w_{tj}|j \in \mathcal{M}\}$ are defined as in the previous scenario.
The parameter $v_{tj} \ge 0$ denotes the attractiveness of resource $j \in \mathcal{M}$ to customer $t$, and $v_{t0} \ge 0$ denotes the attractiveness of the outside option.
If an assortment $\mathcal{S} \subseteq \mathcal{M}$ of available resources is offered to customer $t$, she chooses option $j \in \mathcal{S}^+$ with probability proportional to its attractiveness, that is, with probability ${v_{tj}}/{(\sum_{k \in \mathcal{S}^+} v_{tk})}$. To ensure well-defined choice probabilities, if $\sum_{k \in \mathcal{S}^+} v_{tk} = 0$, we assume that the customer chooses the outside option with probability 1.
Note that the BAM model proposed in \cite{luce1959individual} requires $v_{t0} > 0$. 
Here, we allow $v_{t0} = 0$ so that the accept-or-reject scenario is included as a special case (see Lemma~\ref{lemma:special_case}). 
We also remark that BAM encompasses the widely used multinomial logit choice model \citep{mcfadden1974conditional}.
The decision-maker’s objective is to maximize the total expected revenue over the planning horizon by determining, in each period, which assortment of available resources to offer. 
Again, in Section~\ref{sec:general_arrivals}, we extend the Bernoulli arrivals to \textit{general arrivals}, where $l_t, r_t, \{w_{tj}|j \in \mathcal{M} \}$ and $\{ v_{tj} | j \in \mathcal{M}\}$ can also be random and follow known distributions.

We illustrate the setting using the boutique hotel room-booking example introduced in \cite{rusmevichientong2023revenue}. In this case, a customer arrives randomly in each period, with an exogenously specified itinerary over a time interval. Upon receiving the request, the decision-maker presents an assortment of available rooms, from which the customer selects using a choice model. Like ours, \cite{rusmevichientong2023revenue} assume that the customers' choice models are known; in their empirical studies, they use multinomial logit models, although their methodology applies more broadly to general choice models. 


For simplicity, we continue to use $\mathcal{I}$ to denote an instance and $\overline{V}(\mathcal{I})$ to denote the optimal objective value. Since attractiveness can be zero, the previous scenario is a special case of the BAM-based scenario, as shown below (with the proof available in Appendix~\ref{sec:proof_special_case}):
\begin{lemma}
\label{lemma:special_case}
Under Bernoulli arrivals, any $\alpha$-approximation policy for the BAM-based scenario with $v_{t0} = 0$ immediately yields an $\alpha$-approximation policy for the accept-or-reject scenario.
\end{lemma}



\section{Warm-up: Optimal Policy for Single-Resource Case}
\label{sec:warm_up}
In this section, we present a polynomial-time optimal policy for a special case: the accept-or-reject scenario under Bernoulli arrivals with a single resource $(M=1)$. This illustrates the decomposable property, which forms the foundation of our algorithmic design in more general settings.
 
A straightforward approach to solving this case is through dynamic programming, as described below. Let $\mathcal{I}$ be an instance of the problem. Since there is only one resource, we use $w_t$ to denote $w_{t1}$ for simplicity. We capture the availability status of the resource across time slots with a binary vector $\mathbf{s} \in \{0,1\}^N$, where $s_i = 1$ indicates that slot $i \in \mathcal{N}$ is available. We assume $s_0 = s_{N+1} = 0$ for convenience. Let $\mathbf{1}_{[a,b]}$ denote a binary vector with ones in the interval $[a, b]$ and zeros elsewhere. Denote by $G_t(\mathbf{s})$ the optimal expected revenue to obtain at the beginning of period $t \in \mathcal{T}$ given state $\mathbf{s}$. 
\tw{Then, $\overline{V}(\mathcal{I}) = G_1(\mathbf{1}_{[1,N]})$ can be computed recursively as follows:
\begin{align}
\tag{DP}
\label{eq:naive_dp}
G_t(\mathbf{s}) = \begin{cases}
G_{t+1}(\mathbf{s}) + p_t \max \Bigl\{ w_t + G_{t+1}(\mathbf{s} - \mathbf{1}_{[l_t, r_t]}) - G_{t+1}(\mathbf{s}), 0 \Bigr\}, & \text{if } \mathbf{1}_{[l_t, r_t]} \le \mathbf{s}, \\
G_{t+1}(\mathbf{s}), & \text{otherwise},
\end{cases}
\end{align}
for all $t \in \mathcal{T}$ and $\mathbf{s} \in \{0,1\}^N$, with the boundary condition $G_{T+1}(\mathbf{s}) = 0$ for all $\mathbf{s} \in \{0, 1\}^N$. }
This dynamic program follows the principle that if the requested interval $[l_t, r_t]$ is available under the current status, the decision-maker chooses the option that maximizes the expected future revenue. However, the number of possible states grows exponentially with the number of time slots, making the naive implementation computationally impractical.

To this end, we demonstrate that the dynamic program can be implemented more efficiently by reducing the number of states to a polynomial size. We begin by introducing a definition that characterizes the structure of the resource status:
\begin{definition}[Maximal Sequence] Given any vector $\mathbf{s} \in \{0, 1\}^N$, an interval $[a,b]$ with $1 \le a \le b \le N$ is called a \emph{maximal sequence} in $\mathbf{s}$ (denoted by $[a,b] \sim \mathbf{s}$) if and only if $s_i = 1$ for all $i \in [a,b]$, and $s_{a-1} = 0$ and $s_{b+1} = 0$. \end{definition}
This concept, introduced in \cite{rusmevichientong2023revenue} and \cite{zhu2023assign}, has been used to develop decomposable structures for evaluating policies or deriving optimal algorithms in static seat allocation problems. 
A state $\mathbf{s}$ may contain several maximal sequences.  
The key observation is that future requests interact with each maximal sequence independently: a request $[l_t, r_t]$ affects only the unique maximal sequence containing it (if any).  
This allows us to decompose the value function into per-sequence components.

\tw{
\begin{proposition}[Decomposability]
\label{prop:decomp}
For any instance $\mathcal{I}$ of the accept-or-reject scenario under Bernoulli arrivals with $M = 1$, any state $\mathbf{s}$, and any $t \in \mathcal{T}$,
\begin{align}
\label{eq:decomp}
G_t(\mathbf{s}) = \sum_{[a,b] \sim \mathbf{s}} G_t(\mathbf{1}_{[a,b]}).
\end{align}
\end{proposition}
}
\tw{
\proof{Proof of Proposition \ref{prop:decomp}.}
The proof is by backward induction on $t$.  
At $t = T+1$, both sides of \eqref{eq:decomp} are zero.  
Assume the claim holds for $t+1$.  
If $[l_t,r_t] \nsubseteq [a,b]$ for every maximal sequence $[a,b] \sim \mathbf{s}$, applying \eqref{eq:naive_dp} and \eqref{eq:decomp} gives
\[
G_t(\mathbf{s}) = G_{t+1}(\mathbf{s}) = \sum_{[a,b] \sim \mathbf{s}} G_{t+1}(\mathbf{1}_{[a,b]})
= \sum_{[a,b] \sim \mathbf{s}} G_{t}(\mathbf{1}_{[a,b]}).
\]
If $[l_t,r_t] \subseteq [a',b']$ for exactly one maximal sequence $[a',b']$, let $\mathcal{U} \triangleq \{ [a,b]| [a,b] \sim \mathbf{s} \text{ and } [a,b] \ne [a',b'] \}$ be the full set of maximal sequences except $[a',b']$, and we have 
\begin{align*}
G_t(\mathbf{s})
=& \sum_{[a,b] \in \mathcal{U}} G_{t+1}(\mathbf{1}_{[a,b]}) + G_{t+1}(\mathbf{1}_{[a',b']}) + p_t \max \Bigl\{ w_t + G_{t+1}(\mathbf{1}_{[a',b']} - \mathbf{1}_{[l_t,r_t]}) - G_{t+1}(\mathbf{1}_{[a',b']}), 0 \Bigr\} \\
=& \sum_{[a,b] \in \mathcal{U}} G_{t}(\mathbf{1}_{[a,b]}) + G_{t}(\mathbf{1}_{[a',b']}) = \sum_{[a,b] \sim \mathbf{s}} G_{t}(\mathbf{1}_{[a,b]}),
\end{align*}
where the first equality is obtained by  applying \eqref{eq:decomp} to \eqref{eq:naive_dp}, and the second equality is obtained by using \eqref{eq:naive_dp} again.
This completes the proof.  
\hfill \Halmos
}

Leveraging Proposition~\ref{prop:decomp}, any value \( G_t(\mathbf{s}) \) can be computed in \( O(TN^2) \) time since what we really need to compute is $O(TN^2)$ state values. Therefore, the dynamic programming approach \eqref{eq:naive_dp} admits a polynomial-time implementation. 
\begin{corollary}
\label{corollary:special}
The accept-or-reject scenario under Bernoulli arrivals with a single resource (\(M = 1\)) admits a polynomial-time optimal policy.
\end{corollary}




\section{Accept-or-Reject Scenario}
\label{sec:single_item}
In this section, we present a \((1 - 1/e)\)-approximation policy for the accept-or-reject scenario under Bernoulli arrivals. Motivated by the decomposed dynamic programming approach in the single-resource case (\(M = 1\)) (see Equation~\eqref{eq:decomp}), we formulate a fluid relaxation \eqref{eq:lp_single} for the general case where \(M > 1\). Building on the optimal solution to this relaxation, we design a proposal-discarding algorithm that leverages virtual resource statuses to preserve probabilistic independence across resources.

\subsubsection*{Fluid Relaxation.} 
The first key component of our policy is the fluid relaxation for the online optimum. For any given policy, let \( x_{tj}([a,b]) \) denote the probability that \([a,b]\) is a maximal sequence of resource \( j \in \mathcal{M} \) at the beginning of period \( t \in \mathcal{T} \). Let \( y_{tj}([a,b]) \) denote the joint probability that (i) \([a,b]\) is a maximal sequence of resource \( j \) at the beginning of period \( t \), and (ii) the request arriving in period \( t \) is allocated the time slots \([l_t, r_t] \subseteq [a,b]\) of resource \( j \).
Then, the fluid relaxation is presented as follows:
\begin{align}
\label{eq:lp_single}
\tag{LP}
     \max_{\mathbf{x} \ge \mathbf{0}, \mathbf{y} \ge \mathbf{0}} \quad& \sum_{j \in \mathcal{M}}  \sum_{t \in \mathcal{T}} \sum_{1 \le a \le b \le N} w_{tj} y_{tj}([a,b]) \\
    \label{eq:allocate_single}
    \tag{Online}
    \text{ s.t.} \qquad& y_{tj}([a, b]) \le x_{tj}([a,b]) \cdot p_t, &&  \forall j \in \mathcal{M},  t \in \mathcal{T},  1 \le a \le b \le N \\
    \label{eq:no_allocate}
    \tag{Feasibility}
    \quad& y_{tj}([a, b]) \le \mathbbm{1}\{ [l_t,r_t] \subseteq [a,b] \}, && \forall j \in \mathcal{M},  t \in \mathcal{T},  1 \le a \le b \le N \\
    \quad& x_{tj}([a,b]) = x_{t-1,j}([a,b]) \quad - \quad y_{t-1,j}([a,b])  && +
    \sum_{ r_{t-1} \le b' \le N} \mathbbm{1}\{ b+1 = l_{t-1} \}y_{t-1,j}([a, b']) \notag
    \\
    \label{eq:trans_single}
    \tag{Balance}
    & \quad + \sum_{1 \le a' \le l_{t-1} } \mathbbm{1}\{ r_{t-1} = a-1 \} y_{t-1,j}([a',b]), && \forall j \in \mathcal{M},  2 \le t \le T,  1 \le a \le b \le N  \\
    \label{eq:boundary_single}
    \tag{Boundary}
    \quad& x_{1j}([a,b]) = \mathbbm{1}\{ [a,b]=[1,N]\}, &&  \forall j \in \mathcal{M},  1 \le a \le b \le N \\
    \label{eq:across_single}
    \tag{Capacity}
    \quad& \sum_{j \in \mathcal{M}}\sum_{1 \le a \le b \le N} y_{tj}([a,b]) \le p_t, &&   \forall t \in \mathcal{T}. 
\end{align}
Here, $\mathbbm{1}\{\cdot\}$ is the indicator function, and we use $\texttt{LP}(\mathcal{I})$ to denote the optimal objective value of \eqref{eq:lp_single}. We explain the constraints as follows. 
Inequality \eqref{eq:allocate_single} is due to the fact that since arrivals are independent of resource statuses, the allocation probability for any online policy is bounded by $x_{tj}([a,b]) \cdot p_t$. 
\tw{This constraint, first introduced in online bipartite matching by \cite{torrico2022dynamic} and \cite{papadimitriou2024online}, is crucial for online algorithms, as it does not necessarily hold for the offline optimum.}
Inequality \eqref{eq:no_allocate} says that an allocation is feasible only if $[l_t, r_t] \subseteq [a,b]$. Equation \eqref{eq:trans_single} tracks how maximal sequences evolve across periods, either remaining unchanged or being split by allocations.  Equation \eqref{eq:boundary_single} initializes the resource status with full availability. Inequality \eqref{eq:across_single} ensures that the total allocation probability across resources does not exceed the request arrival probability.

This fluid relaxation generalizes the decomposed dynamic program \eqref{eq:decomp} to the case of multiple resources. When $M=1$, constraints \eqref{eq:allocate_single}--\eqref{eq:boundary_single} exactly correspond to the dynamic program. The additional constraint \eqref{eq:across_single} enforces that, in the multi-resource setting, each request is fulfilled at most once.
Since any feasible policy induces, in expectation, a feasible solution to this linear program, the following upper bound on the optimal value follows immediately (a simple proof is provided in Appendix~\ref{subsec:proof_in_reject_or_accept}). 

\begin{lemma} 
\label{lemma:upper} For any instance $\mathcal{I}$ of the accept-or-reject scenario under Bernoulli arrivals, we have $\overline{V}(\mathcal{I}) \le \emph{\texttt{LP}}(\mathcal{I})$. 
\end{lemma}

\subsubsection*{Virtual Resource Status.}
We generalize the proposal-discarding framework \citep{braverman2025new} by introducing a key concept, \textit{virtual resource status}.
To motivate this introduction, we take \eqref{eq:lp_single} as an example and first explain the challenge of translating the optimal solution of the fluid relaxation into an actionable policy.
Specifically, in \eqref{eq:lp_single}, the ratio \({y_{tj}([a,b])}/{[x_{tj}([a,b]) \cdot p_t]}\) can be interpreted as the intended conditional probability of allocating the interval \([l_t, r_t]\) of resource \(j\) to request \(t\), given that \([a,b]\) is its current maximal sequence and request $t$ arrives. However, these probabilities may not be compatible across resources. For example, consider the case where \(p_t = 1\), and for two resources \(j = 1, 2\), we have \(x_{tj}([a_j, b_j]) = y_{tj}([a_j, b_j]) = 0.5\). 
Although this satisfies constraints \eqref{eq:allocate_single} and \eqref{eq:across_single}, strictly adhering to the conditional probabilities would imply that both resources allocate time slots simultaneously, which may not be feasible unless the events that \([a_j, b_j] \) is the maximal sequence of resource $j$ for \(j = 1, 2\) are disjoint. Enforcing such a negative correlation, however, is nontrivial. 

\tw{To address this issue, \cite{braverman2025new} introduce a proposal-discarding framework for the online matching problem, which operates in two stages.
In the proposal stage, each available offline node generates a proposal according to the probabilities prescribed by the fluid relaxation, regardless of whether the online node will actually arrive (following the idea introduced by \citealt{braverman2022max}). 
In the allocation stage, the algorithm selects the max-weight proposal to match with the arriving online node, while all unselected proposers are discarded independently, as if they had also been used for matching. 
\cite{braverman2025new} show that if the proposals are also independent, then this mechanism preserves probabilistic independence across offline nodes while ensuring consistency with the marginal behavior specified by the fluid relaxation.
We adopt this proposal-discarding framework and further generalize the discarding idea by introducing the virtual resource status, which is deemed available only when the corresponding actual resource status is available. 
This virtual status facilitates keeping probabilistic independence across resources and the marginal probabilities from the fluid relaxation. 
}


\subsubsection*{Algorithm Design.}
We now describe our policy. At each period $t \in \mathcal{T}$, given the virtual status $\bar{\mathbf{s}}^j_t \in \{0,1\}^N$ of each resource $j \in \mathcal{M}$ (carried over from the previous period), the policy proceeds in two stages: proposal and allocation.

\paragraph{Proposal stage.}  
This stage occurs before the realization of the request at time $t$. Each resource independently submits a proposal to allocate its time slots, regardless of whether the request arrives. For each \( j \in \mathcal{M} \), let \( [a,b] \) be the maximal sequence in \( \bar{\mathbf{s}}^j_t \) such that \( [a,b] \supseteq [l_t, r_t] \), if such a sequence exists. Then, resource \( j \) is included in the proposal set \( \mathcal{P} \) with probability \( {y_{tj}([a,b])}/{[x_{tj}([a,b]) \cdot p_t]} \). 
\tw{This step is inspired by the independent proposal mechanism in \cite{braverman2022max}, where each offline node in the online matching problem independently submits a proposal if it is unmatched. 
Our proposal step instead operates on a virtual status, following the approach of \cite{braverman2025new}, and we further extend it from a binary setting to one that incorporates maximal sequences.}

\paragraph{Allocation stage.} 
Among all proposed resources, the policy selects the one with the highest revenue:
\(
j^* = \arg\max_{j \in \mathcal{P}} \{ w_{tj} \}.
\)
If the request arrives, the policy allocates the interval $[l_t, r_t]$ from resource $j^*$, and updates its virtual status by setting $\bar{\mathbf{s}}^{j^*}_{t+1} \leftarrow \bar{\mathbf{s}}^{j^*}_t - \mathbf{1}_{[l_t, r_t]}$. If the request does not arrive, the virtual status of resource $j^*$ is carried forward.
For each $j \in \mathcal{P} \setminus \{ j^* \}$, the virtual status is updated as follows: with probability $p_t$, set $\bar{\mathbf{s}}^j_{t+1} \leftarrow \bar{\mathbf{s}}^j_t - \mathbf{1}_{[l_t, r_t]}$; with probability $1 - p_t$, keep $\bar{\mathbf{s}}^j_{t+1} = \bar{\mathbf{s}}^j_t$. For all resources not in $\mathcal{P}$, the virtual status is simply carried forward to the next period. 
\tw{The selection of the highest-revenue proposal follows the same approach as \cite{braverman2022max}. 
The step of updating the virtual resource status for $\mathcal{P} \backslash \{ j^* \}$ is inspired by the discarding idea in \cite{braverman2025new}, where each offline node that submits a proposal is discarded with the arrival probability of the online node.
}

The full procedure is summarized in Algorithm~\ref{alg:single_item}.

\begin{algorithm}
\caption{$(1-1/e)$-Approximation Policy for the Accept-or-Reject Scenario under Bernoulli Arrivals} 
\label{alg:single_item}
\begin{algorithmic}[1]
\REQUIRE Instance $\mathcal{I} = \{\mathcal{M}, \mathcal{N}, \{ \theta_t\}_{t \in \mathcal{T}} \}$
\STATE Initialize $\bar{\mathbf{s}}^j_1 \gets \mathbf{1}_{[1,N]}$ for all $j \in \mathcal{M}$. Let $(\mathbf{x}, \mathbf{y})$ be the optimal solution to \eqref{eq:lp_single}.
\FOR{each period $t \in \mathcal{T}$}
    \STATE \textbf{Proposal Stage:} Initialize $\mathcal{P} \gets \emptyset$.
    \COMMENT{Before the realization of request $t$'s arrival}
    \begin{ALC@g}
    \STATE \textbf{for} each $j \in \mathcal{M}$ with $[a,b] \sim \bar{\mathbf{s}}^j_t$ s.t. $[l_t, r_t] \subseteq [a,b]$ \textbf{do}: Add $j$ to $\mathcal{P}$ w.p. ${y_{tj}([a,b])}/{[x_{tj}([a,b])\cdot p_t]}$ independently.
    \end{ALC@g}

    \STATE \textbf{Allocation Stage:} Initialize $\bar{\mathbf{s}}^j_{t+1} \gets \bar{\mathbf{s}}^j_t$ for all $j \in \mathcal{M}$.
    \begin{ALC@g}
    \STATE $j^* = \arg\max_{j \in \mathcal{P}} w_{tj}$ or $0$ if $\mathcal{P} = \emptyset$.
    \IF{$\mathcal{P} \ne \emptyset$ and request $t$ arrives}
    \STATE Allocate $[l_t, r_t]$ of resource $j^*$ to $t$. Update $\bar{\mathbf{s}}^{j^*}_{t+1} \gets \bar{\mathbf{s}}^{j^*}_t - \mathbf{1}_{[l_t, r_t]}$.
    \ENDIF
    \STATE \textbf{for} each $j \in \mathcal{P} \setminus \{j^*\}$ \textbf{do}: Update $\bar{\mathbf{s}}^j_{t+1} \gets \bar{\mathbf{s}}^j_t - \mathbf{1}_{[l_t, r_t]}$ w.p. $p_t$ independently.
    \end{ALC@g}
\ENDFOR
\end{algorithmic}
\end{algorithm}

Next, we show the key properties for the virtual resource status.
Let $\mathbf{s}^j_t \in \{0,1\}^N$ denote the actual availability status of resource $j$ at beginning of period $t$. 
Then, the constructed virtual resource status in Algorithm~\ref{alg:single_item} will satisfy the following properties, which we will establish shortly:
\begin{property}
\label{property:virtual}
\begin{enumerate}
    \item[$(i)$] [{\sc Lower Bound}] \(\bar{\mathbf{s}}^j_t \le \mathbf{s}^j_t \) almost surely for all \( t \in \mathcal{T}, j \in \mathcal{M} \).
    \item[$(ii)$] [{\sc Marginal Probability}] \(\Pr\{ [a,b] \sim \bar{\mathbf{s}}^j_t \} = x_{tj}([a,b]) \text{ for all } t \in \mathcal{T}, j \in \mathcal{M}, 1\le a\le b \le N\).
    \item[$(iii)$] [{\sc Independence}] The collections \( \{ \bar{\mathbf{s}}^j_t \}_{t \in \mathcal{T}} \) are  independent across resources $j \in \mathcal{M}$.
\end{enumerate}
\end{property}

These properties are central to both the design and analysis of our algorithm. (i) ensures that the virtual status is a conservative approximation of the true availability, thereby guaranteeing the feasibility of allocations (see line 8 in Algorithm~\ref{alg:single_item}). (ii) guarantees alignment with the marginal distributions prescribed by the fluid relaxation. (iii) enables a tractable implementation by preserving probabilistic independence across resources. 
Now we formally state that our constructed virtual resource statuses indeed satisfy Property~\ref{property:virtual} and defer the proof that proceeds by induction on $t$ to Appendix~\ref{subsec:proof_in_reject_or_accept}. 
\begin{proposition}
\label{prop:virtual_proerty}
The virtual resource statuses \(\{ \bar{\mathbf{s}}^j_t \}_{j \in \mathcal{M}, t \in \mathcal{T}}\) in Algorithm~\ref{alg:single_item} satisfy Property~\ref{property:virtual}.
\end{proposition}

\subsubsection*{Performance Analysis.}
We show that Algorithm~\ref{alg:single_item} achieves a \((1 - 1/e)\)-approximation guarantee. 

\begin{theorem}\label{thm:1 - 1/e}
The accept-or-reject scenario under Bernoulli arrivals admits a \((1 - 1/e)\)-approximation policy, given by Algorithm~\ref{alg:single_item}.
\end{theorem}

\tw{At a high level, our proof shows that in each period, because (i) the proposals are independent across resources, (ii) their total probabilities do not exceed one, and (iii) we select the highest-revenue proposal, we can directly apply the correlation gap for submodular functions to obtain a lower bound on the expected revenue.
}


\proof{Proof of Theorem \ref{thm:1 - 1/e}.}
Let $R_t \triangleq \max_{j \in \mathcal{P}} \{ w_{tj} \}$ denote the revenue in period $t$, conditioned on the event that customer $t$ arrives.
First, the proposal probability for resource $j$ in period $t$ is:
\begin{align}
\label{eq:proposal_prob}
\Pr[j \in \mathcal{P}] = \sum_{[a,b] \supseteq [l_t,r_t]} x_{tj}([a,b]) \cdot \frac{y_{tj}([a,b])}{x_{tj}([a,b])\cdot p_t}
= \sum_{[a,b] \supseteq [l_t, r_t]} \frac{y_{tj}([a,b])}{p_t}
= \frac{1}{p_t} \sum_{1\le a \le b \le N} y_{tj}([a,b]),
\end{align}
where the first equality follows from line 4 in Algorithm~\ref{alg:single_item} and Property~\ref{property:virtual}(ii), and the last equality holds because the summation is extended over all intervals \( [a,b] \) such that $1\le a \le b \le N$, with additional terms contributing zero by constraint \eqref{eq:no_allocate}.
By Constraint~\eqref{eq:across_single} of the linear program, the total proposal probability in period $t$ satisfies:
$$
\sum_{j \in \mathcal{M}} \Pr[j \in \mathcal{P}] = \frac{1}{p_t} \sum_{j \in \mathcal{M}} \sum_{1\le a\le b \le N} y_{tj}([a,b]) \le 1.
$$

\tw{Next, we use the result that bounds the correlation gap of monotone submodular functions.
\begin{lemma}[\citealt{agrawal2010correlation}]
\label{lemma:correlation_gap}
Let $f: 2^{[n]} \to \mathbb{R}_{+}$ be a nonnegative monotone submodular function with $f(\emptyset) = 0$. 
Fix any vector $\mathbf{z} = (z_1,\ldots,z_n) \in [0,1]^n$ and consider two random sets $\mathcal{S}$ and $\mathcal{I}$ over $[n]$ such that:
(i) $\mathcal{S}$ has an arbitrary joint distribution with marginals $\Pr[j \in \mathcal{S}] = z_j$ for all $j \in [n]$;
(ii) $\mathcal{I}$ is generated by including each element $j$ independently with probability $z_j$.
Then
\[
\mathbb{E}\bigl[f(\mathcal{I})\bigr] 
\;\ge\; \left(1 - \frac{1}{e}\right) \, \mathbb{E}\bigl[f(\mathcal{S})\bigr].
\]
\end{lemma}
}

\tw{To apply Lemma~\ref{lemma:correlation_gap}, let $n = M$, let $\mathcal{I} = \mathcal{P}$, and define 
$f(\mathcal{X}) = \max_{j \in \mathcal{X}} w_{tj}$ for any $\mathcal{X} \subseteq [n]$. 
Let $\mathcal{S}$ be a random set that contains at most one element, where 
$\Pr[\mathcal{S} = \{j\}] = \Pr[j \in \mathcal{P}]$ for all $j$, and $\Pr[\mathcal{S} = \emptyset] = 1 - \sum_j \Pr[j \in \mathcal{P}]$.
Then Lemma~\ref{lemma:correlation_gap} implies that
\begin{align}
\label{eq:lower_bound_revenue_period}
\mathbb{E}[R_t] 
= \mathbb{E}[f(\mathcal{I})] 
\ge \left(1 - \frac{1}{e}\right) \mathbb{E}[f(\mathcal{S})] 
= \left(1 - \frac{1}{e}\right) \frac{\sum_{j \in \mathcal{M}} \sum_{a,b} w_{tj} \, y_{tj}([a,b])}{p_t}.
\end{align}
}
Summing over all periods, we conclude:
\[
V^\pi(\mathcal{I}) = \sum_{t \in \mathcal{T}} p_t\cdot \mathbb{E}[R_t] 
\ge \left(1 - \frac{1}{e} \right) \sum_{t,j,a,b} w_{tj} y_{tj}([a,b]) 
= \left(1 - \frac{1}{e} \right) \texttt{LP}(\mathcal{I}) \ge (1-\frac{1}{e}) \overline{V}({\mathcal{I}}),
\]
where the first equality follows from the fact that the customer $t$'s arrival is independent of $R_t$, the second equality follows from the fact that $(\mathbf{x},\mathbf{y})$ is the optimal solution to \eqref{eq:lp_single}, and the last inequality is due to Lemma \ref{lemma:upper}. \hfill \Halmos

\subsubsection*{Hardness Results.}
We complement our approximation guarantees with several hardness results.

\tw{
\paragraph{Inapproximability.}
The accept-or-reject scenario under Bernoulli arrivals contains a special case in which the maximum interval length \(L\) is at most one. 
This case has been shown to be PSPACE-hard to approximate within a certain constant strictly less than one (though the bound exceeds \(0.99\)) \citep{papadimitriou2024online, braverman2025new}. 
For larger values of \(L\), we prove a stronger hardness result for the offline version, which immediately transfers to the online setting:
}

\tw{
\begin{proposition}
\label{prop:hardness_accept_reject}
Given any \( \epsilon > 0 \), it is NP-hard to approximate the accept-or-reject scenario under Bernoulli arrivals within a factor of \( 0.95 + \epsilon \), even when all arrival probabilities satisfy \( p_t = 1 \).
\end{proposition}
}

\tw{The proof appears in Section~\ref{sec:hardness}. 
It follows a reduction from 3-SAT similar to that of \cite{arkin1987scheduling}, except that we employ the inapproximability result of 3-SAT and construct a gap-preserving reduction.
}

\tw{
\paragraph{Tightness.}
\citet[Section~3.5]{braverman2022max} present a simple example showing that their policy cannot achieve an approximation ratio better than \(1 - 1/e\). 
On this example, Algorithm~\ref{alg:single_item} coincides with their policy; therefore, Algorithm~\ref{alg:single_item} also cannot achieve an approximation ratio exceeding \(1 - 1/e\), establishing tightness. 
Moreover, for the offline version of the accept-or-reject scenario, the best known approximation ratio is also \(1 - 1/e\) \citep{bhatia2007algorithmic}\footnote{\cite{bhatia2007algorithmic} consider the case where \(w_{tj} = w_t\) is uniform across \(j\), but their algorithm easily extends to heterogeneous weights via a max-weight selection step.}. 
Thus, any improvement in our approximation ratio would likely require progress on the offline problem first; even then, transferring such an improvement to the online setting remains challenging. 
We leave this as a direction for future work.
}

\tw{
\paragraph{Integrality gap.}
\cite{papadimitriou2024online} show that the integrality gap of \eqref{eq:lp_single} is at most \(1-\frac{1}{2e} \approx 0.816\) when \(L = 1\). 
We extend their example to obtain a slightly stronger upper bound when \(L = 2\) (see Appendix~\ref{sec:integrality_gap}):
\begin{proposition}
\label{prop:integrality_gap_accept_reject}
The integrality gap of the fluid relaxation \eqref{eq:lp_single} in the accept-or-reject scenario under Bernoulli arrivals is at most \(0.8134\); that is,
\[
\inf_{\mathcal{I}} \left \{ \frac{\overline{V}(\mathcal{I})}{\emph{\texttt{LP}}(\mathcal{I})} \right \} \le 0.8134.
\]
\end{proposition}
}

\section{BAM-Based Scenario}
\label{sec:assortment}
In this section, we present a 
\minor{$0.271$}-approximation policy for the BAM-based scenario under Bernoulli arrivals. This policy builds on Algorithm~\ref{alg:single_item} for the accept-or-reject scenario by continuing to utilize virtual resource statuses. Unlike the previous case, however, the decision-maker must now offer an assortment of resources from which the customer selects using a choice model. To handle this added complexity, we incorporate new coupling techniques that maintain probabilistic independence across resources while enabling effective assortment offerings.

\subsubsection*{Sales-Based Fluid Relaxation.} 
We adopt the sales-based linear programming (SBLP) framework of \cite{gallego2015general}, which captures customer choice behavior under the basic attraction model without explicitly enumerating all feasible assortments. For any policy, we define the following decision variables:
(i) $x_{tj}([a,b])$: the probability that $[a,b]$ is a maximal sequence of resource $j$ at the beginning of period $t$;
(ii) $y_{tj}([a,b])$: the joint probability that resource $j$ is selected by customer $t$ (implying that $j$ is in the offered assortment), and $[a,b] \supseteq [l_t,r_t]$ is a maximal sequence of $j$;
(iii) $y^0_{tj}([a,b])$: the joint probability that resource $j$ is included in the assortment, customer $t$ chooses the outside option, and $[a,b] \supseteq [l_t,r_t]$ is a maximal sequence of $j$;
(iv) $y^0_t$: the probability that customer $t$ arrives and chooses the outside option (aggregated over all assortments).
The SBLP formulation is given as:
\begin{align}
\label{eq:sblp}
\tag{SBLP}
    \max_{\mathbf{x} \ge \mathbf{0}, \mathbf{y} \ge \mathbf{0}} \quad& \sum_{j \in \mathcal{M}}  \sum_{t \in \mathcal{T}} \sum_{1 \le a \le b \le N} w_{tj} y_{tj}([a,b]) \\
     \tag{Online}
    \label{eq:allocate_assortment}
    \text{ s.t.}   \qquad& 
    y^0_{tj}([a,b]) + y_{tj}([a, b]) \le x_{tj}([a,b]) \cdot p_t, && \forall j \in \mathcal{M},  t \in \mathcal{T},  1 \le a \le b \le N \\
    \tag{Feasibility}
    \label{eq:no_allocate_assortment}
     \quad& y^0_{tj}([a,b]) + y_{tj}([a, b]) \le \mathbbm{1}\{[l_t, r_t] \subseteq [a,b], v_{tj} > 0  \}, &&  \forall j \in \mathcal{M},  t \in \mathcal{T},  1 \le a \le b \le N \\
    \quad& x_{tj}([a,b]) = x_{t-1,j}([a,b]) \quad\quad  - \quad \quad y_{t-1,j}([a,b])  && +
    \sum_{ r_{t-1} \le b' \le N} \mathbbm{1}\{ b+1 = l_{t-1} \}y_{t-1,j}([a, b']) \notag
    \\
    \label{eq:balance}
    \tag{Balance}
    &  \qquad\qquad + \sum_{1 \le a' \le l_{t-1} } \mathbbm{1}\{ r_{t-1} = a-1 \} y_{t-1,j}([a',b]), && \forall j \in \mathcal{M},  2 \le t \le T,  1 \le a \le b \le N  \\
    \tag{Boundary}
    \label{eq:boundary_assortment}
    \quad& x_{1j}([a,b]) = \mathbbm{1}\{ [a,b]=[1,N]\}, && \forall j \in \mathcal{M},  1 \le a \le b \le N \\
    \tag{Capacity}
    \label{eq:capacity}
    \quad& y^0_t + \sum_{j \in \mathcal{M}}\sum_{1 \le a \le b \le N} y_{tj}([a,b]) = p_t, &&  \forall t \in \mathcal{T} \\
    &
    \tag{Scale}
    \label{eq:scale}
     v_{t0}y_{tj}([a,b]) = v_{tj}y^0_{tj}([a,b]), && \forall j \in \mathcal{M}, t  \in \mathcal{T}, 1 \le a \le b \le N\\
    \tag{Opt-out}
    \label{eq:no_purchase}
    \quad&  \sum_{1 \le a \le b \le N} y^0_{tj}([a,b]) \le y^0_t, && \forall j \in \mathcal{M}, t \in \mathcal{T}.
\end{align}
We now provide explanations for each constraint. Constraint  
\eqref{eq:allocate_assortment} bounds the joint probability that customer $t$ sees resource $j$ and 
$[a,b]$ is its maximal available sequence, by the product of the availability probability and the customer arrival probability.
Constraint \eqref{eq:no_allocate_assortment} excludes infeasible allocations where the interval is not contained in the maximal sequence or the resource has zero attractiveness (since it will not affect the customer's choice).
Equations \eqref{eq:balance} and \eqref{eq:boundary_assortment} track the evolution of maximal sequences and initialize full availability, respectively.
Equation \eqref{eq:capacity} ensures the total probability of selection (including the outside option) equals the arrival probability.
Equation \eqref{eq:scale} enforces proportional choice under the basic attraction model.
Constraint \eqref{eq:no_purchase} says that the aggregated probability of choosing the outside option with resource $j$ being offered cannot exceed the total non-selection probability.
Finally, the optimal value of this LP (denoted by $\texttt{SBLP}(\mathcal{I})$) provides an upper bound on the expected revenue of any online policy (a straightforward proof is provided in Appendix~\ref{sec:appendix_assortment}):
\begin{lemma}
\label{lemma:upper_assortment}
For any instance $\mathcal{I}$ of the BAM-based scenario under Bernoulli arrivals, 
\(
\overline{V}(\mathcal{I}) \le \emph{\texttt{SBLP}}(\mathcal{I}).
\)
\end{lemma}

\subsubsection*{Algorithm Design.}
Our policy for the BAM-based scenario also builds on the concept of virtual resource status, but introduces additional complexity to preserve probabilistic independence when offering assortments. Let $\mathbf{s}^j_t$ and $\bar{\mathbf{s}}^j_t$ denote the actual and virtual availability statuses, respectively, of resource $j \in \mathcal{M}$ at the beginning of period $t \in \mathcal{T}$. Again, the policy proceeds in two stages: the proposal stage and the assortment recommendation stage.

\paragraph{Proposal stage.} 
Similar to Algorithm~\ref{alg:single_item}, at the beginning of period $t \in \mathcal{T}$, each resource $j \in \mathcal{M}$ independently submits a proposal with probability ${[y^0_{tj}([a,b]) + y_{tj}([a,b])]}/{[x_{tj}([a,b]) \cdot p_t]}$, where $[a,b] \supseteq [l_t, r_t]$ is the maximal sequence in the virtual resource status $\bar{\mathbf{s}}^j_t$ (we take \(\tfrac{0}{0} = 0\) ). This indicates a willingness to be included in the assortment, regardless of whether the customer arrives. If no such interval $[a,b]$ exists in $\bar{\mathbf{s}}^j_t$, resource $j$ does not submit a proposal.

\tw{
\paragraph{Assortment recommendation stage.}
Let $\mathcal{P}$ denote the set of resources that submitted proposals. 
To determine the assortment $\mathcal{S}$ to recommend, 
we select the subset of $\mathcal{P}$ that yields the highest expected revenue, i.e.,
\begin{align}
\label{eq:s_from_p}
    \mathcal{S} 
    \triangleq 
    \arg\max_{\mathcal{S}' \subseteq \mathcal{P}}
    \left\{
        \frac{\sum_{j \in \mathcal{S}'} w_{tj} v_{tj}}
        {v_{t0} + \sum_{j \in \mathcal{S}'} v_{tj}}
    \right\}.
\end{align}
This step can be viewed as a generalization of selecting the highest-revenue proposal in the accept-or-reject scenario, since when $v_{t0} = 0$ it reduces exactly to that earlier step.}
To update the virtual resource status, we randomly generate a subset of resources $\mathcal{Q}$, where the virtual status of each resource $j \in \mathcal{Q}$ is updated as $\bar{\mathbf{s}}^j_{t+1} \gets \bar{\mathbf{s}}^j_t - \mathbf{1}_{[l_t, r_t]}$. For the remaining resources in $\mathcal{M} \setminus \mathcal{Q}$, the virtual status is carried forward unchanged.
To maintain Property~\ref{property:virtual}, the set $\mathcal{Q}$ is determined using the sampling function $\textsc{Random}(\mathbf{q}, \mathbf{q}', j^*)$, which randomly selects a subset of $\mathcal{M}$. 
Here, $\mathbf{q}, \mathbf{q}' \in [0,1]^M$ are probability vectors, and $j^*$ denotes the customer's choice. The choice of $\mathbf{q}$ and $\mathbf{q}'$ will be detailed later.
\tw{The \textsc{Random} function serves as a technical component to generalize the idea of independently discarding unselected proposers. 
In the accept-or-reject scenario, all proposers become unavailable (either by being discarded or matched) independently, and this distribution is coupled with the allocation decision. 
We extend this idea so that the coupling now deals with the customer's random choice.}
The function $\textsc{Random}$ is implemented via a Markov chain coupling procedure, presented in Appendix~\ref{subsec:random}, and summarized as follows. 

\begin{proposition}
\label{prop:random}
    \begin{itemize}
        \item[$(i)$] {\sc [Complexity]} $\emph{\textsc{Random}}(\mathbf{q}, \mathbf{q}', \tilde{j})$ runs in $O(M)$ time.
        \item[$(ii)$] {\sc [Inclusion]} If $\tilde{j} \in \mathcal{M}$, then $\tilde{j} \in \emph{\textsc{Random}}(\mathbf{q}, \mathbf{q}', \tilde{j})$ almost surely.
        \item[$(iii)$] {\sc [Independence]} Fix any vectors $\mathbf{q}, \mathbf{q}' \in [0,1]^M$ such that $\sum_{j \in \mathcal{M}} q'_j \le 1$ and 
        \begin{align}
        \label{eq:regular_condtion}
              \frac{q'_j}{ 1 - \sum_{j' = j+1}^M q'_{j'}  } \le q_j, \quad \forall j \in \mathcal{M}.
        \end{align} 
         Suppose the input of  index $\tilde{j}$ is random with the distribution $P$ that is defined by $\mathbf{q}'$
        \begin{align}
        \label{eq:distribution}
            P(j) \triangleq \begin{cases}
                q'_j , &\text{ \emph{if} } j \in \mathcal{M} \\
                1 - \sum_{j' \in \mathcal{M}} q'_{j'}, & \text{ \emph{if} } j = 0.
            \end{cases}
        \end{align}
        Then, conditioned on $\mathbf{q}$ and $\mathbf{q}'$, the random output of $\emph{\textsc{Random}}(\mathbf{q}, \mathbf{q}', \tilde{j})$ has the following  distribution:
        \begin{align}
            {\Pr}_{\tilde{j} \sim P }\Big[  \emph{\textsc{Random}}(\mathbf{q}, \mathbf{q}',\tilde{j}) =\mathcal{X} \Big| \mathbf{q}, \mathbf{q}' \Big] = \prod_{j \in \mathcal{X}} q_j \prod_{j \in \mathcal{M}\backslash \mathcal{X}} (1-q_j), \qquad \forall \ \mathcal{X} \subseteq \mathcal{M}.
        \end{align}
    \end{itemize}
\end{proposition}

In our policy, to generate the set $\mathcal{Q}$ of resources with the virtual resource status to be updated, given the proposal set $\mathcal{P}$ and the assortment $\mathcal{S}$, we specify $\mathbf{q}$ and $\mathbf{q}'$ as follows:
\begin{align}
\label{eq:q}
    \mathbf{q} = [q_j]_{j \in \mathcal{M}}, & \text{ where } q_j = \mathbbm{1}\{ j \in \mathcal{P}\} \cdot p_t \cdot \frac{v_{tj}}{v_{t0} + v_{tj}}, \\
\label{eq:q'}
    \mathbf{q}' = [q'_j]_{j \in \mathcal{M}}, & \text{ where } q'_j = \mathbbm{1}\{ j \in \mathcal{S} \} \cdot p_t \cdot \frac{v_{tj}}{v_{t0} + \sum_{j' \in \mathcal{S}}v_{tj'}}.
\end{align}
Let $j^*$ denote the customer $t$'s choice, or the outside option $0$ if customer $t$ does not arrive. Then, we set $\mathcal{Q}$ as ${\textsc{Random}}(\mathbf{q}, \mathbf{q}', j^*)$. Our policy is presented in Algorithm~\ref{alg:assortment_based}.

\begin{algorithm}
            \caption{\minor{$0.271$}-Approximation Policy for the BAM-Based Scenario under Bernoulli Arrivals} 
            \label{alg:assortment_based}
            \begin{algorithmic}[1]
            \REQUIRE Instance $\mathcal{I} \triangleq \{\mathcal{M}, \mathcal{N}, \{ \theta_t\}_{t \in \mathcal{T}} \}$.
            \STATE Initialize $\bar{\mathbf{s}}^j_1 \gets \mathbf{1}_{[1,N]}$ for all $j \in \mathcal{M}$. Let $(\mathbf{x}, \mathbf{y})$ be the optimal solution to \eqref{eq:sblp}.
            \FOR{each period $t \in \mathcal{T}$}
            \STATE \textbf{Proposal Stage:} Initialize $\mathcal{P} \gets \emptyset$.
            \COMMENT{Before the realization of request $t$'s arrival}
            \begin{ALC@g}
            \STATE \textbf{for} each $j \in \mathcal{M}$ with $[a,b] \sim \bar{\mathbf{s}}^j_t$ s.t. $[l_t, r_t] \subseteq [a,b]$ \textbf{do}: Add $j$ to $\mathcal{P}$ w.p. ${[y^0_{tj}([a,b]) + y_{tj}([a,b])]}/{[x_{tj}([a,b])\cdot p_t]}$ independently.
            \end{ALC@g}
            \STATE \textbf{Assortment Recommendation Stage:} Initialize $\bar{\mathbf{s}}^j_{t+1} \gets \bar{\mathbf{s}}^j_t$ for all $j \in \mathcal{M}$.
            \begin{ALC@g}
            \STATE Determine the  assortment $\mathcal{S}$ by \eqref{eq:s_from_p}.
            \STATE \textbf{if} customer $t$ arrives \textbf{then}: Offer $\mathcal{S}$ and let $j^*$ denote her choice; \textbf{else}: $j^* \gets 0$.
            \STATE $\mathcal{Q} \leftarrow \textsc{Random}(\mathbf{q}, \mathbf{q}', j^*)$ with $\mathbf{q}$ and $\mathbf{q}'$ defined by \eqref{eq:q} and \eqref{eq:q'}, respectively.
            \STATE \textbf{for} each $j \in \mathcal{Q}$ \textbf{do}:  $\bar{\mathbf{s}}^j_{t+1} \gets \bar{\mathbf{s}}^j_t - \mathbf{1}_{[l_t, r_t]}$.
            \end{ALC@g}
            \ENDFOR
            \end{algorithmic} 
        \end{algorithm}

We show that the virtual resource statuses we constructed indeed satisfy Property~\ref{property:virtual} as follows.
\begin{proposition}
\label{prop:virtual_property_assortment}
The virtual resource statuses $\{ \bar{\mathbf{s}}^j_t \}_{j \in \mathcal{M}, t \in \mathcal{T}}$ generated by Algorithm~\ref{alg:assortment_based} satisfy Property~\ref{property:virtual}. Moreover, for each period $t \in \mathcal{T}$, the candidate assortment $\mathcal{S}$ is feasible to offer, in the sense that $\mathcal{S} \subseteq \{ j \in \mathcal{M} \mid \mathbf{1}_{[l_t, r_t]} \le \mathbf{s}^j_t \}$.
\end{proposition}

The proof proceeds by induction on the period \( t \in \mathcal{T} \), with the key step relying on Proposition~\ref{prop:random}.  
In period \( t \), it is straightforward to verify that, given an assortment \( \mathcal{S} \) and the vectors \( \mathbf{q}, \mathbf{q}' \) defined in \eqref{eq:q} and \eqref{eq:q'}, then \( j^* \) follows the distribution \( P \) in~\eqref{eq:distribution}.  
Moreover, \( \mathbf{q} \) and \( \mathbf{q}' \) satisfy the regularity condition~\eqref{eq:regular_condtion}, since
\[
\frac{q'_j}{\,1 - \sum_{j' = j+1}^M q'_{j'}\,}
= 
\frac{\mathbbm{1}\{ j \in \mathcal{S} \}\, p_t v_{tj}}
     {v_{t0} + \sum_{j' \in \mathcal{S}} v_{tj'} - \sum_{j' = j+1}^M \mathbbm{1}\{ j' \in \mathcal{S} \}\, p_t v_{tj'}}
\le 
\mathbbm{1}\{ j \in \mathcal{S} \}\, \frac{p_t v_{tj}}{v_{t0} + v_{tj}}
\le q_j.
\]
The first equality follows from the definition of \( q'_j \) in~\eqref{eq:q'}.  
For the first inequality, if \( j \notin \mathcal{S} \), the inequality is immediate; and if \( j \in \mathcal{S} \), it holds because the denominator on the right-hand side is no more than that on the left.  
The final inequality uses the fact that \( \mathcal{S} \subseteq \mathcal{P} \).  
A complete proof of Proposition~\ref{prop:virtual_property_assortment} is provided in Appendix~\ref{subsec:proofs_assortment}.

\subsubsection*{Performance Analysis.} 
Now we establish the performance guarantee for Algorithm~\ref{alg:assortment_based}. 

\tw{
\begin{theorem} \label{thm:1/8} 
    The BAM-based scenario under Bernoulli arrivals admits a \minor{$0.271$}-approximation policy, given by Algorithm~\ref{alg:assortment_based}.
\end{theorem}
}

\tw{
\proof{Proof of Theorem \ref{thm:1/8}.}
The proof analyzes the expected revenue obtained in a single period \(t\), conditioned on a customer arrival. 
At a high level, we partition the resources into two groups based on their attractiveness and derive lower bounds on the expected revenue from using only resources from each group. 
We then combine these bounds to obtain the final guarantee.
}


\paragraph{{Preliminaries.}}
Fix any period $t \in \mathcal{T}$.
Since $y_{tj}([a,b]) = y^0_{tj}([a,b]) = 0$ for any resource $j$ with $v_{tj} = 0$ (by Constraint~\eqref{eq:no_allocate_assortment}), we restrict attention to the set
$
\mathcal{M}_t \triangleq \{ j \in \mathcal{M} \mid v_{tj} > 0 \}
$
in period $t$.
For any $j \in \mathcal{M}_t$, the probability of submitting a proposal is:
\begin{align}
\label{eq:prob_proposal}
\begin{split}
    & \Pr\{ j \in \mathcal{P} \} = \sum_{1\le a \le b \le N: [a,b] \supseteq [l_t,r_t]}\Pr\{ [a,b] \sim \bar{\mathbf{s}}^j_t \text{ and } j \in \mathcal{P} \} \\
    & \ = \sum_{1\le a \le b \le N:[a,b]\supseteq[l_t,r_t]} x_{tj}([a,b]) \cdot \frac{y^0_{tj}([a,b]) + y_{tj}([a,b])}{x_{tj}([a,b])\cdot p_t}  = \frac{\sum_{1 \le a \le b \le N} y_{tj}([a,b])}{p_t} \cdot \frac{v_{t0} + v_{tj}}{ v_{tj}}, 
\end{split}
\end{align}
where the second equality uses the marginal property of virtual resource status (Property~\ref{property:virtual}) and line 4 in Algorithm~\ref{alg:assortment_based}, and the third equality follows from Constraint~\eqref{eq:scale} and by extending the summation so that it is over all intervals $[a,b]$ with $1\le a \le b \le N$, where the additional terms contribute zero due to Constraint~\eqref{eq:no_allocate_assortment}. For ease of notation, we use $y_j$ to denote $\sum_{1 \le a \le b \le N} y_{tj}([a,b])$.

Next, we partition $\mathcal{M}_t$ based on resource attractiveness.
\minor{Fix a threshold parameter $\rho>0$, and set $\lambda \triangleq 1+1/\rho$.}
Let $\mathcal{M}^{\texttt{L}} \triangleq \{j \in \mathcal{M}_t| \minor{v_{tj} > \rho v_{t0}} \}$ and $\mathcal{M}^{\texttt{S}} \triangleq \{j \in \mathcal{M}_t| \minor{v_{tj} \le \rho v_{t0}} \}$ be a partition of $\mathcal{M}_t$ by the size of the attractiveness of a resource to the customer. 
In addition, define 
\[L \triangleq \Bigl(\sum_{j \in \mathcal{M}^{\texttt{L}}} \sum_{1\le a\le b \le N} y_{tj}([a,b]) \Bigr)/p_t = \Big(\sum_{j \in \mathcal{M}^{\texttt{L}}} y_j \Big)/p_t \text{ and } S \triangleq \Bigl(\sum_{j \in \mathcal{M}^{\texttt{S}}} \sum_{1\le a\le b \le N} y_{tj}([a,b]) \Bigr) / p_t = \Big(\sum_{j \in \mathcal{M}^{\texttt{S}}} y_j \Big)/p_t,
\]
which represent the scaled expected allocations to large- and small-attractiveness resources, respectively.
In the following, we analyze the expected revenue contributed by 
$\mathcal{M}^{\texttt{S}}$ and $\mathcal{M}^{\texttt{L}}$ separately, 
and then combine the results to obtain the overall bound.

\tw{
\paragraph{Bounding the expected revenue from $\mathcal{M}^{\emph{\texttt{S}}}$.}
Let $\mathcal{P}^{\texttt{S}} \triangleq \{ j \in \mathcal{P} \mid \minor{v_{tj} \le \rho v_{t0}} \}$.
Conditioned on the arrival of customer $t$, we lower bound the expected revenue from offering assortment $\mathcal{P}^{\texttt{S}}$ as follows:
\begin{align}
\label{eq:lower_bound_small}
\begin{split}
   & \mathbb{E} \left[ \frac{\sum_{j \in \mathcal{P}^{\texttt{S}}} w_{tj} v_{tj}}
{v_{t0} + \sum_{j \in \mathcal{P}^{\texttt{S}}} v_{tj}} \right] = \sum_{j \in \mathcal{M}^{\texttt{S}}} \Pr\{j \in \mathcal{P}^{\texttt{S}} \} \cdot \mathbb{E}\left[ \frac{w_{tj}v_{tj}}{v_{t0} + \sum_{j' \in \mathcal{P}^{\texttt{S}}} v_{tj'}} \middle| j \in \mathcal{P}^{\texttt{S}} \right] \\
& \ge \sum_{j \in \mathcal{M}^{\texttt{S}}} \Pr\{j \in \mathcal{P}^{\texttt{S}} \} \cdot \mathbb{E}\left[ \frac{w_{tj}v_{tj}}{v_{t0} + v_{tj} + \sum_{j' \in \mathcal{M}^{\texttt{S}}: j' \ne j} \Pr \{j' \in \mathcal{P}^{\texttt{S}} \} v_{tj'}} \middle| j \in \mathcal{P}^{\texttt{S}} \right] \\
& \ge \sum_{j \in \mathcal{M}^{\texttt{S}}} \Pr\{j \in \mathcal{P}^{\texttt{S}} \} \cdot  \frac{w_{tj}v_{tj}}{\minor{v_{t0} + v_{tj} + (1+\rho)Sv_{t0}}} \ge \minor{\frac{1}{(1+\rho)S+1}} \sum_{j \in \mathcal{M}^{\texttt{S}}} \Pr\{j \in \mathcal{P}^{\texttt{S}} \} \cdot  \frac{w_{tj}v_{tj}}{v_{t0} + v_{tj} } \\
& = \minor{\frac{1}{(1+\rho)S+1}} \sum_{j \in \mathcal{M}^{\texttt{S}}} \frac{w_{tj}y_j}{p_t}.
\end{split}
\end{align}
Above, the first inequality follows from the inequality from convexity, say 
$
\mathbb{E}[{x'}/{(x'' + X)}] \ge {x'}/{(x''+\mathbb{E}[X])}
$ for any $x',x'' > 0$ and any nonnegative random variable $X$, and the fact that events $j \in \mathcal{P}^{\texttt{S}}$ are independent across resources $j \in \mathcal{M}^{\texttt{S}}$ (by Line 4 in Algorithm~\ref{alg:assortment_based}).
The second inequality uses the fact that 
\begin{align*}
\minor{\sum_{j \in \mathcal{M}^{\texttt{S}}} \Pr \{j \in \mathcal{P}^{\texttt{S}} \}  v_{tj}}
&\minor{=\sum_{j \in \mathcal{M}^{\texttt{S}}} \frac{y_j (v_{t0} + v_{tj})}{p_t}
\le (1+\rho) v_{t0}\sum_{j \in \mathcal{M}^{\texttt{S}}} \frac{y_j}{p_t}
= (1+\rho)Sv_{t0},}
\end{align*}
where the first equality uses \eqref{eq:prob_proposal}, the inequality uses \minor{$v_{tj} \le \rho v_{t0}$} for $j \in \mathcal{M}^{\texttt{S}}$, and the last equality is by the definition of $S$. The third inequality in \eqref{eq:lower_bound_small} follows from
\[
\minor{\begin{aligned}
v_{t0}+v_{tj}+(1+\rho)Sv_{t0}
&\le \big((1+\rho)S+1 \big)(v_{t0}+v_{tj}).
\end{aligned}}
\]
The last equality uses \eqref{eq:prob_proposal}.
}

\tw{
\paragraph{Bounding the expected revenue from $\mathcal{M}^{\emph{\texttt{L}}}$.}
Throughout this step, we assume $L > 0$.
Let $\mathcal{P}^{\texttt{L}}$ be the singleton set from $\mathcal{P}\cap\mathcal{M}^{\texttt{L}}$ with the highest revenue, i.e., 
\[
\mathcal{P}^{\texttt{L}} \triangleq \Big\{ \arg\max_{j \in \mathcal{P} \cap \mathcal{M}^{\texttt{L}}}
\Bigl\{
\dfrac{w_{tj} v_{tj}}{v_{t0} + v_{tj}}
\Bigr\} \Big\}.
\]
For ease of our analysis, for any value $w \ge 0$, we denote by $\mathcal{S}_w \triangleq \{j\in \mathcal{M}^{\texttt{L}}| w\le {w_{tj} v_{tj}}/{(v_{t0} + v_{tj})} \}$. Then, conditioned on the customer arrival, we lower bound the expected revenue from offering $\mathcal{P}^{\texttt{L}}$ as follows:
\begin{align}
\label{eq:lower_bound_large}
\begin{split}
   & \mathbb{E}\left[\frac{\sum_{j \in \mathcal{P}^{\texttt{L}}} w_{tj} v_{tj}}{v_{t0} + \sum_{j \in \mathcal{P}^{\texttt{L}}} v_{tj}} \right] = \int_0^\infty \Pr \Big\{ \exists j \in \mathcal{P}\cap \mathcal{M}^{\texttt{L}} \text{ such that } w \le \frac{w_{tj}v_{tj}}{v_{t0} + v_{tj}} \Big\} \mathrm{d} w \\
    & = \int_{0}^\infty 1 - \prod_{j \in \mathcal{S}_w} (1 - \Pr\{j \in \mathcal{P} \}) \mathrm{d}w 
    = \int_{0}^\infty 1 - \prod_{j \in \mathcal{S}_w} (1 - \frac{y_j}{p_t}\cdot \frac{v_{t0}+v_{tj}}{v_{tj}}) \mathrm{d}w \\
    & \ge \int_{0}^\infty 1 - \Big(1 - (\sum_{j \in \mathcal{S}_w } \frac{y_j}{p_t}\cdot \frac{v_{t0} +v_{tj}}{v_{tj}})/|\mathcal{S}_w| \Big)^{|\mathcal{S}_w|}\mathrm{d}w \ge \int_{0}^\infty 1 - \exp( - \sum_{j \in \mathcal{S}_w } \frac{y_j}{p_t}\cdot \frac{v_{t0} +v_{tj}}{v_{tj}}) \mathrm{d}w \\
    & \ge \minor{\frac{1-e^{-\lambda L}}{\lambda L}} \int_0^\infty \sum_{j \in \mathcal{S}_w } \frac{y_j}{p_t}\cdot \frac{v_{t0} +v_{tj}}{v_{tj}} \mathrm{d}w = \minor{\frac{1-e^{-\lambda L}}{\lambda L}} \sum_{j \in \mathcal{M}^{\texttt{L}}} \frac{w_{tj}y_j}{p_t}.
\end{split}
\end{align}
Above, the first equality follows from the definition of $\mathcal{P}^{\texttt{L}}$. The second equality holds since the events $j \in \mathcal{P}$ are independent across resources $j \in \mathcal{M}$. The third equality uses \eqref{eq:prob_proposal}. The first inequality uses the fact that ${y_j}/{p_t}\cdot {(v_{t0}+v_{tj})}/{v_{tj}} = \Pr\{ j \in \mathcal{P}\} \le 1$ and the AM-GM inequality that $\prod_{i \in [n]} x_i \le ({\sum_{i \in [n]} x_i}/{n})^{n}$ for any $n \in \mathbb{N}$ and nonnegative numbers $x_1,\dots,x_n$. The second inequality  uses the fact that $(1-x)^{1/x} \le 1/e$ for any $0< x \le 1$ and the assumption that $L > 0$. The last inequality uses the inequality
\begin{align*}
    \minor{\sum_{j \in \mathcal{S}_w } \frac{y_j}{p_t}\cdot \frac{v_{t0} +v_{tj}}{v_{tj}} \stackrel{(\text{by } v_{t0} < v_{tj}/\rho) }{\le} \lambda \sum_{j \in \mathcal{S}_w } \frac{y_j}{p_t} \le \lambda L}
\end{align*}
and the concavity of $y = 1-e^{-x}$. The last equality in \eqref{eq:lower_bound_large} is by the definition of $\mathcal{S}_w$.
}

\tw{
\paragraph{Putting everything together.}
Since the recommended assortment $\mathcal{S}$ is a subset of $\mathcal{P}$ with the highest expected revenue (as in \eqref{eq:s_from_p}), its expected revenue is higher than those of $\mathcal{P}^{\texttt{S}}$ and $\mathcal{P}^{\texttt{L}}$. We consider two cases depending on the value of $L$.
If $L = 0$, just by \eqref{eq:lower_bound_small}  and $S \le 1$, we get that 
\[
\mathbb{E} \left[\frac{\sum_{j\in \mathcal{S}} w_{tj} v_{tj}}{v_{t0} + \sum_{j \in \mathcal{S}} v_{tj}} \right] \ge \mathbb{E} \left[ \frac{\sum_{j \in \mathcal{P}^{\texttt{S}}} w_{tj} v_{tj}}
{v_{t0} + \sum_{j \in \mathcal{P}^{\texttt{S}}} v_{tj}} \right] \ge \minor{\frac{1}{2+\rho}} \sum_{j \in \mathcal{M}^{\texttt{S}}} \frac{w_{tj} y_j}{p_t} = \minor{\frac{1}{2+\rho}} \sum_{j \in \mathcal{M}} \frac{w_{tj} y_j}{p_t}.
\]
}

\tw{If $L>0$, combining with lower bounds from \eqref{eq:lower_bound_small} and \eqref{eq:lower_bound_large}, we get that 
\begin{align}
\label{eq:lower_bound_S}
\begin{split}
    & \mathbb{E} \left[\frac{\sum_{j\in \mathcal{S}} w_{tj} v_{tj}}{v_{t0} + \sum_{j \in \mathcal{S}} v_{tj}} \right] 
    \ge \mathbb{E} \left[ \max \middle\{ \frac{\sum_{j\in \mathcal{P}^{\texttt{S}}} w_{tj} v_{tj}}{v_{t0} + \sum_{j \in \mathcal{P}^{\texttt{S}}} v_{tj}}, \frac{\sum_{j\in \mathcal{P}^{\texttt{L}}} w_{tj} v_{tj}}{v_{t0} + \sum_{j \in \mathcal{P}^{\texttt{L}}} v_{tj}} \middle\} \right] \\
    & \ge \max \left\{ \mathbb{E} \left[ \frac{\sum_{j \in \mathcal{P}^{\texttt{S}}} w_{tj} v_{tj}}
{v_{t0} + \sum_{j \in \mathcal{P}^{\texttt{S}}} v_{tj}} \right], \mathbb{E} \left[ \frac{\sum_{j \in \mathcal{P}^{\texttt{L}}} w_{tj} v_{tj}}
{v_{t0} + \sum_{j \in \mathcal{P}^{\texttt{L}}} v_{tj}} \right] \right\} \\
& \ge \max \left\{ \minor{\frac{1}{(1+\rho)S+1}} \sum_{j \in \mathcal{M}^{\texttt{S}}} \frac{w_{tj}y_j}{p_t}  , \minor{\frac{1-e^{-\lambda L}}{\lambda L}} \sum_{j \in \mathcal{M}^{\texttt{L}}} \frac{w_{tj}y_j}{p_t} \right\} \\
& \ge \minor{\frac{1}{\max\left\{3+\rho,\ 1+\frac{\lambda}{1-e^{-\lambda}}\right\}}} \sum_{j \in \mathcal{M}} \frac{w_{tj}y_j}{p_t},
\end{split}
\end{align}
where the last inequality follows from the fact that $L+S\le 1$  and the following technical lemma (see the proof in Appendix~\ref{subsec:proof_technical_lemma}):
\begin{lemma}
\label{lemma:technical}
    \minor{For any $\rho>0$ and $\lambda=1+1/\rho$, given any values of $S,X,Y \ge 0$ and $L > 0$ with $X+Y=1$ and $L+S\le 1$, we have 
    \[
    \max\left\{ \frac{X}{(1+\rho)S+1},\frac{(1-e^{-\lambda L})Y}{\lambda L} \right\} \ge \frac{1}{\max\left\{3+\rho,\ 1+\frac{\lambda}{1-e^{-\lambda}}\right\}}.
    \]}
\end{lemma}
}

\minor{The lower bound above is optimized by choosing $\rho$ to minimize $\max\{3+\rho,\ 1+\lambda/(1-e^{-\lambda})\}$ with $\lambda=1+1/\rho$. Since $3+\rho$ is increasing in $\rho$, while $1+\lambda/(1-e^{-\lambda})$ is decreasing in $\rho$, the optimal value is attained when the two terms are equal. Solving this equation gives $\rho^*\approx 0.6867$ and $\lambda^*=1+1/\rho^*\approx 2.4563$, yielding
\[
\frac{1}{3+\rho^*}=\frac{1}{1+\lambda^*/(1-e^{-\lambda^*})}\approx 0.2712.
\]}
Therefore, combining the two cases above, the expected revenue for period $t$ is at least \minor{$0.271 \sum_{j \in \mathcal{M}} w_{tj} y_j$}. Then, summing over different periods $t \in \mathcal{T}$, we get that 
\[
\minor{V^{\pi}(\mathcal{I})  \ge 0.271 \sum_{t \in \mathcal{T}} \sum_{j \in \mathcal{M}} \sum_{1\le a\le b\le N}w_{tj} y_{tj}([a,b]) = 0.271 \texttt{SBLP}(\mathcal{I}) \ge 0.271 \overline{V}(\mathcal{I}).} \hfill \Halmos
\]

\tw{
\begin{remark}
    Our analysis in period \(t\) relies on the fact that each resource enters \(\mathcal{P}\) independently, and we aim to lower bound the expected revenue obtained by offering the highest-revenue assortment \(\mathcal{S} \subseteq \mathcal{P}\). 
It remains an intriguing question whether the bound can be tightened, either by improving the bounds for the small- or large-attractiveness groups separately (e.g., through other choices for \(\mathcal{P}^{\texttt{S}}\) and \(\mathcal{P}^{\texttt{L}}\)), or by developing a unified analysis that avoids partitioning resources into these two groups. 
\end{remark}
}

\tw{
\subsubsection*{Integrality Gap.} Finally, we establish an upper bound on the integrality gap of \eqref{eq:sblp} using a family of single-slot instances (\(L = 1\)). 
\begin{proposition}
\label{prop:integrality_gap_BAM}
The integrality gap of the fluid relaxation \eqref{eq:sblp} in the BAM-based scenario under Bernoulli arrivals is at most \(3/4\); that is, 
\[
\inf_{\mathcal{I}} \left\{ \frac{\overline{V}(\mathcal{I})}{ \emph{\texttt{SBLP}}(\mathcal{I}) } \right\} \le \frac{3}{4}.
\]
\end{proposition}
}

\tw{
\section{Extension to General Arrivals}
\label{sec:general_arrivals}
In this section, we extend our constant-factor algorithm for the BAM-based scenario with Bernoulli arrivals to a more general setting where the request type in each period is also random. 
We then show that, under general arrivals, the resulting algorithm achieves a \((1 - 1/e)^2\)-approximation ratio for the accept-or-reject scenario and a \minor{\(0.271(1 - 1/e)\)}-approximation ratio for the BAM-based scenario.
}

\tw{
\subsubsection*{Model.}
We generalize the Bernoulli arrivals in the BAM-based scenario for period \(t\) to
\begin{align}
\label{eq:request_general}
    \theta_t = \Big\{ \big(p^{(k)}_t,\, l^{(k)}_t,\, r^{(k)}_t,\, \{ w^{(k)}_{tj} \mid j \in \mathcal{M} \}, \{ v^{(k)}_{tj} \mid j \in \mathcal{M}^+ \}\big) \Big\}_{k=1}^{K_t},
\end{align}
where \(K_t\) is the number of possible request types in period \(t\), and the arrival probabilities satisfy \(\sum_{k=1}^{K_t} p^{(k)}_t = 1\) without loss of generality. 
In each period \(t\), the arriving request is of type \(k \in [K_t]\) with probability \(p^{(k)}_t\). 
If the request is of type \(k\), it demands interval \([l^{(k)}_t, r^{(k)}_t]\), chooses resource \(j\) with probability 
\(
v^{(k)}_{tj}/(v^{(k)}_{t0} + \sum_{j' \in \mathcal{S}} v^{(k)}_{tj'})
\) when offered assortment \(\mathcal{S}\), and generates revenue \(w^{(k)}_{tj}\) if it chooses resource $j$.
}

\tw{
The accept-or-reject scenario under general arrivals corresponds to \eqref{eq:request_general} without the choice parameters \(\{v^{(k)}_{tj}\}\). 
That is, in period $t$, for a customer of type \(k\), the decision-maker may allocate interval \([l^{(k)}_t, r^{(k)}_t]\) of any available resource \(j\), yielding revenue \(w^{(k)}_{tj}\).
Analogous to the equivalence established in Lemma~\ref{lemma:special_case}, we obtain the following result (see Appendix~\ref{sec:proof_special_case} for proof):
}

\tw{
\begin{lemma}
\label{lemma:special_case_general}
Under general arrivals, any \(\alpha\)-approximation policy for the BAM-based scenario with \(v^{(k)}_{t0} = 0\) immediately yields an \(\alpha\)-approximation policy for the accept-or-reject scenario.
\end{lemma}
}

\tw{Therefore, establishing a \((1-1/e)^2\)-approximation ratio for the special case \(v^{(k)}_{t0} = 0\) immediately implies the same guarantee for the accept-or-reject scenario under general arrivals.
}

\tw{
\subsubsection*{Fluid Relaxation.}
First, we formulate a linear program analogous to~\eqref{eq:sblp} by extending the variable set \(\mathbf{y}\) to incorporate multiple request types (see Appendix~\ref{appendix:full_lp_general} for the full details):
\begin{align}
\label{eq:sblp_general}
\tag{SBLP-General}
     \max_{\mathbf{x} \ge \mathbf{0}, \mathbf{y} \ge \mathbf{0}} \quad& \sum_{j \in \mathcal{M}}  \sum_{t \in \mathcal{T}} \sum_{1 \le a \le b \le N} \sum_{k=1}^{K_t} w^{(k)}_{tj} y^{(k)}_{tj}([a,b]) \\
    \label{eq:allocate_general}
    \tag{Online}
    \text{ s.t.} \qquad& y^{(k)0}_{tj}([a,b]) +  y^{(k)}_{tj}([a, b]) \le x_{tj}([a,b]) \cdot p^{(k)}_t, &&  \forall j \in \mathcal{M},  t \in \mathcal{T},  1 \le a \le b \le N, k \in [K_t] \\
    \notag
    & \eqref{eq:no_allocate_general}, \eqref{eq:trans_general}, \eqref{eq:boundary_general}, \\
    \notag
    & \eqref{eq:capacity_general}, \eqref{eq:scale_general}, \eqref{eq:no_purchase_general}. 
\end{align}
Here, \( x_{tj}([a,b]) \) retains the same interpretation as in \eqref{eq:sblp}, while \( y^{(k)}_{tj}([a,b]) \) denotes the joint probability that \([a,b] \supseteq [l^{(k)}_t, r^{(k)}_t]\) is a maximal sequence of resource \( j \) at the beginning of period \( t \), and resource $j$ is selected by customer $t$ of type $k$. In addition, $y^{(k)0}_{tj}([a,b])$ denotes the joint probability that resource $j$ is recommended to customer $t$ of type $k$, the customer chooses the outside option, and $[a,b] \supseteq [l^{(k)}_t,r^{(k)}_t]$ is the maximal sequence of resource $j$. 
With slight abuse of notation, we continue to use \( \overline{V}(\mathcal{I}) \) and \( \texttt{SBLP}(\mathcal{I}) \) to denote the online optimum and the optimal objective value of \eqref{eq:sblp_general}, respectively.  
Since the expected decisions under any feasible policy yield a feasible solution to \eqref{eq:sblp_general}, we obtain the following upper bound:
}

\tw{
\begin{lemma} 
\label{lemma:upper_general} 
For any instance \( \mathcal{I} \) of the BAM-based scenario under general arrivals,  we have
\(
\overline{V}(\mathcal{I}) \le \emph{\texttt{SBLP}}(\mathcal{I}).
\)
\end{lemma}
}

\tw{
\subsubsection*{Algorithm Design.}
We now extend Algorithm~\ref{alg:assortment_based} to handle general arrivals.  
During the {proposal stage}, we construct a proposal set for each potential request type.  
Specifically, for type $k$ in period \( t \), we independently add each resource \( j \) to its proposal set \( \mathcal{P}^{(k)} \) with probability 
\(
\big(y^{(k)0}_{tj}([a,b]) + y^{(k)}_{tj}([a,b])\big)\big/\big(x_{tj}([a,b]) \cdot p^{(k)}_t \big)
\)
given that \( [a,b] \sim \bar{\mathbf{s}}^j_t \) (Line 7 in Algorithm~\ref{alg:assortment_general}).
}

\tw{In the allocation stage, we can no longer directly recommend, for each type \( k \), the assortment from \( \mathcal{P}^{(k)} \) with the highest expected revenue. This is because the randomness in the request type makes it impossible to preserve both the customer-choice distribution and the probabilistic independence of virtual-resource statuses across resources. 
To overcome this difficulty, we design a coupling procedure that incurs an additional multiplicative factor of \(1 - 1/e\) in performance. When combined with the \(1 - 1/e\) (resp., \minor{\(0.271\)}) factor obtained under Bernoulli arrivals, this yields an overall approximation ratio of \((1 - 1/e)^2\) (resp., \minor{\(0.271(1 - 1/e)\)}).}

\tw{In particular, analogous to \eqref{eq:s_from_p}, let $\mathcal{S}^{(k)}$ denote the revenue-maximizing assortment drawn from $\mathcal{P}^{(k)}$, i.e.,
\begin{align}
\label{eq:s_from_p_general}
    \mathcal{S}^{(k)} 
    \triangleq 
    \arg\max_{\mathcal{S}' \subseteq \mathcal{P}^{(k)}}
    \left\{
        \frac{\sum_{j \in \mathcal{S}'} w^{(k)}_{tj} v^{(k)}_{tj}}
        {v^{(k)}_{t0} + \sum_{j \in \mathcal{S}'} v^{(k)}_{tj}}
    \right\}.
\end{align}
Let $\phi_1,\dots,\phi_{K_t}$ be a permutation of $[K_t]$ satisfying
\begin{align}
\label{eq:dfn_r}
    R_t^{(\phi_1)} \ge \dots \ge R_t^{(\phi_{K_t})}, 
    \qquad 
    \text{where } 
    R_t^{(k)} \triangleq
    \frac{\sum_{j \in \mathcal{S}^{(k)}} w^{(k)}_{tj} v^{(k)}_{tj}}
         {v^{(k)}_{t0} + \sum_{j \in \mathcal{S}^{(k)}} v^{(k)}_{tj}}.
\end{align}
Here, \(R_t^{(k)}\) represents the expected revenue from offering assortment \(\mathcal{S}^{(k)}\) to a type-\(k\) customer in period $t$.
Additionally, define
\begin{align}
\label{eq:dfn_gamma}
    \gamma_i = \prod_{i'=1}^{i-1} \bigl(1 - p^{(\phi_{i'})}_t\bigr),
    \qquad \forall i \in [K_t].
\end{align}
Then, as in Line~12 of Algorithm~\ref{alg:assortment_general}, if customer \(t\) is realized to be of type \(k^*\), we recommend the assortment $\mathcal{S}^{(k^*)}$ with probability \(\gamma_i\), where \(\phi_i = k^*\).
}

\tw{To update the virtual resource status, let \(j^*\) denote the resource chosen by the customer, and set \(j^* = 0\) if she selects no resource. 
We then invoke the coupling subroutine
\(
\textsc{General-Random}\bigl(t, k^*, j^*, \boldsymbol{\phi}, \mathbf{Q}, \mathbf{Q}'\bigr),
\)
which outputs a vector \(\mathbf{k} = [k_j]_{j \in \mathcal{M}}\) (see Line~13).  
Here, \(k_j\) indicates that the virtual status of resource \(j\) should transition as if it had been chosen by a type-\(k_j\) customer in period \(t\).  
Accordingly, we update the virtual resource status as in Line~14.
}

\tw{The procedure \textsc{General-Random} may be viewed as a two-dimensional generalization of the subroutine \textsc{Random} used in Algorithm~\ref{alg:assortment_based}: it couples the realized customer type and choice with independent transitions across resources (i.e., generalizing the independent discarding idea to incorporate both type and choice) according to the matrices \(\mathbf{Q}\) and \(\mathbf{Q}'\).
}

\tw{These matrices are defined as follows:
\begin{align}
\label{eq:dfn_Q}
    &\mathbf{Q} = [q_{kj}]_{k \in [K_t],\, j \in \mathcal{M}}, 
    && \text{ where } q_{kj} = p^{(k)}_t \cdot 
    \frac{\mathbbm{1}[j \in \mathcal{P}^{(k)}] \, v^{(k)}_{tj}}
         {v^{(k)}_{t0} + v^{(k)}_{tj}}, \\
\label{eq:dfn_Q'}
    &\mathbf{Q}' = [q'_{kj}]_{k \in [K_t],\, j \in \mathcal{M}}, 
    && \text{ where } q'_{kj} = \gamma_i \cdot p^{(k)}_t \cdot
    \frac{\mathbbm{1}[j \in \mathcal{S}^{(k)}] \, v^{(k)}_{tj}}
         {v^{(k)}_{t0} + \sum_{j' \in \mathcal{S}^{(k)}} v^{(k)}_{tj'}}
    \quad \text{if } \phi_i = k.
\end{align}
Here, \(q_{kj}\) represents the intended probability that a type-\(k\) customer chooses resource \(j\), while \(q'_{kj}\) represents the actual probability under our policy.  
}

\tw{The properties of the coupling procedure are summarized below and proved in Appendix~\ref{subsec:general_random}.
}

\tw{
\begin{proposition}
\label{prop:general_random}
Fix any \( t \in [T] \), any permutation \( \boldsymbol{\phi} \) of \([K_t]\), and any matrices \(\mathbf{Q}, \mathbf{Q}' \in [0,1]^{K_t \times M}\) such that
\[
\sum_{k \in [K_t], j \in \mathcal{M}} q'_{kj} \le 1,
\qquad
\sum_{k \in [K_t]} q_{kj} \le 1 \ \forall j \in \mathcal{M},
\]
and
\begin{align}
\label{eq:regular_condtion_general}
    \frac{q'_{\phi_i,j}}
    {1 - \sum_{i' < i} \sum_{j' \in \mathcal{M}} q'_{\phi_{i'}, j'} 
       - \sum_{j' < j} q'_{\phi_i, j'}}
    \;\le\; q_{\phi_i,j}, 
    \quad \forall i \in [K_t], \ j \in \mathcal{M}.
\end{align}
Suppose the random inputs \((\tilde{k}, \tilde{j})\) satisfy
\begin{align}
\label{eq:distribution_general}
    \Pr\{\tilde{k} = k,\, \tilde{j} = j\} = q'_{kj},
    \quad \forall k \in [K_t],\, j \in \mathcal{M}.
\end{align}
Then \(\textsc{General-Random}(t, \tilde{k}, \tilde{j}, \boldsymbol{\phi}, \mathbf{Q}, \mathbf{Q}')\) satisfies:
\begin{itemize}
    \item[$(i)$] \textsc{[Complexity]}  
    The procedure runs in \(O(MK_t)\) time.
    \item[$(ii)$] \textsc{[Feasibility]}  
    If \(\tilde{j} \in \mathcal{M}\), then \(k_{\tilde{j}} = \tilde{k}\).
    \item[$(iii)$] \textsc{[Independence]}  
    Conditioned on \((t, \boldsymbol{\phi}, \mathbf{Q}, \mathbf{Q}')\), the output \(\mathbf{k}\) satisfies
    \begin{align}
    \label{eq:indendent_general}
        \Pr[\mathbf{k} = \mathbf{k}' \mid t, \boldsymbol{\phi}, \mathbf{Q}, \mathbf{Q}']
        =
        \prod_{j \in \mathcal{M}} q_{k'_j, j},
        \quad
        \forall \mathbf{k}' \in ([K_t]^+)^M,
    \end{align}
    where \(q_{0j} \triangleq 1 - \sum_{k \in [K_t]} q_{kj}\) and $[K_t]^+ \triangleq \{ 0 \} \cup [K_t]$.
\end{itemize}
\end{proposition}
}

\tw{Inequality~\eqref{eq:regular_condtion_general} is the natural two-dimensional analogue of~\eqref{eq:regular_condtion}.  
It can be satisfied because the algorithm introduces an additional submission probability \(\gamma_i\) for offering the assortment \(\mathcal{S}^{(\phi_i)}\) to type-\(\phi_i\) customers.  
That is, if $\mathbf{Q}$ and $\mathbf{Q}'$ are defined by \eqref{eq:dfn_Q} and \eqref{eq:dfn_Q'}, then for any $i \in [K_t]$ and $j \in \mathcal{M}$,
\begin{align}
\label{eq:satisfy_general_random_condition}
\begin{split}
   & \frac{q'_{\phi_i,j}}{
        1 - \sum_{i' < i} \sum_{j' \in \mathcal{M}} q'_{\phi_{i'},j'}
          - \sum_{j' < j} q'_{\phi_i,j'}}
    \le 
    \frac{q'_{\phi_i,j}}{
        1 - \sum_{i' < i} \gamma_{i'} p^{(\phi_{i'})}_t
          - \sum_{j' < j} q'_{\phi_i,j'} }   \\
    &= 
    \frac{q'_{\phi_i,j}}{
        \gamma_i - \sum_{j' < j} q'_{\phi_i,j'} }    
    =
    \frac{\mathbbm{1}[j \in \mathcal{S}^{(\phi_i)}] p^{(\phi_i)}_t
          v^{(\phi_i)}_{tj}}
        { v^{(\phi_i)}_{t0}
          + \sum_{j' \in \mathcal{S}^{(\phi_i)}} v^{(\phi_i)}_{tj'}
          - \sum_{j' < j} 
              \mathbbm{1}[j' \in \mathcal{S}^{(\phi_i)}] p^{(\phi_i)}_t
              v^{(\phi_i)}_{tj'} }     \\
    &\le 
    \frac{\mathbbm{1}[j \in \mathcal{S}^{(\phi_i)}] 
              p^{(\phi_i)}_t v^{(\phi_i)}_{tj}}
        { v^{(\phi_i)}_{t0} + v^{(\phi_i)}_{tj} }
        \le q_{\phi_i,j},
\end{split}
\end{align}
where the first inequality uses $\sum_{j' \in \mathcal{M}} q'_{\phi_{i'},j'} \le \gamma_{i'} p^{(\phi_{i'})}_t$ (from~\eqref{eq:dfn_Q'}),  
the first equality uses $\gamma_i = 1 - \sum_{i'<i} \gamma_{i'} p^{(\phi_{i'})}_t$ (from the definition of $\gamma_i$ in~\eqref{eq:dfn_gamma}),
the second equality substitutes the expression for $q'_{kj}$ in~\eqref{eq:dfn_Q'},  
and the last inequality follows from~\eqref{eq:dfn_Q} and the fact that 
$\mathcal{S}^{(k)} \subseteq \mathcal{P}^{(k)}$.
Consequently, Property~\ref{property:virtual} continues to hold for the virtual resource statuses \(\{ \bar{\mathbf{s}}^j_t \}\) used in Algorithm~\ref{alg:assortment_general}, as shown below (see the proof in Appendix~\ref{subsec:general_arrival_proof}).
\begin{proposition}
\label{prop:virtual_proerty_general}
The virtual resource statuses \(\{ \bar{\mathbf{s}}^j_t \}_{j \in \mathcal{M},\, t \in \mathcal{T}}\) in Algorithm~\ref{alg:assortment_general} satisfy Property~\ref{property:virtual}.  
Moreover, each assortment \(\mathcal{S}^{(k)}\) is feasible for a type-\(k\) request in the sense that
\(
\mathcal{S}^{(k)} 
\subseteq 
\{ j \in \mathcal{M} \mid \mathbf{1}_{[l^{(k)}_t,\, r^{(k)}_t]} \le \mathbf{s}^j_t \}.
\)
\end{proposition}
}



\begin{algorithm}
\caption{\minor{$0.271(1-1/e)$}-Approximation Policy for the BAM-Based Scenario under General Arrivals} 
\label{alg:assortment_general}
\begin{algorithmic}[1]
\REQUIRE Instance $\mathcal{I} = \{\mathcal{M}, \mathcal{N}, \{ \theta_t \}_{t \in \mathcal{T}} \}$
\STATE Initialize $\bar{\mathbf{s}}^j_1 \gets \mathbf{1}_{[1,N]}$ for all $j \in \mathcal{M}$. 
 Let $(\mathbf{x}, \mathbf{y})$ be the optimal solution to \eqref{eq:sblp_general}.
\FOR{each period $t \in \mathcal{T}$}
    \STATE \textbf{Proposal Stage:} \COMMENT{Executed before the realization of request $t$'s arrival}
    \begin{ALC@g}
    \STATE Initialize $\mathcal{P}^{(k)} \gets \emptyset$ for all $k \in [K_t]$.
    \FOR{each request type $k \in [K_t]$}
        \FOR{each resource $j \in \mathcal{M}$ with $[a,b] \sim \bar{\mathbf{s}}^j_t$ and $[l^{(k)}_t, r^{(k)}_t] \subseteq [a,b]$} 
            \STATE Independently add $j$ to $\mathcal{P}^{(k)}$ with probability ${[y^{(k)0}_{tj}([a,b]) + y^{(k)}_{tj}([a,b])]}/{[x_{tj}([a,b]) \cdot p^{(k)}_t]}$.
        \ENDFOR
    \ENDFOR
    \end{ALC@g}

    \STATE \textbf{Allocation Stage:}
    \begin{ALC@g}
    \STATE Initialize $\bar{\mathbf{s}}^j_{t+1} \gets \bar{\mathbf{s}}^j_t$ for all $j \in \mathcal{M}$.
    \STATE Determine $\mathcal{S}^{(k)}$ according to \eqref{eq:s_from_p_general} for each $k \in [K_t]$.
    \IF{a request of type $k^*$ arrives}
        \STATE With probability $\gamma_i$ (where $\phi_i = k^*$), we offer the assortment $\mathcal{S}^{(k^*)}$ and let $j^*$ denote the customer’s choice; otherwise, set $j^* \gets 0$.
    \ENDIF
        \STATE  $\mathbf{k} = [k_j]_{j \in \mathcal{M}} \leftarrow \textsc{General-Random}(t, k^*, j^*, \boldsymbol{\phi}, \mathbf{Q}, \mathbf{Q}')$ with $\boldsymbol{\phi}, \mathbf{Q}$ and $\mathbf{Q}'$ defined by \eqref{eq:dfn_r}, \eqref{eq:dfn_Q} and \eqref{eq:dfn_Q'}.
        \STATE \textbf{for} each $j \in \mathcal{M}$ \textbf{do}: 
        \textbf{ if } $k_j \ne 0$ \textbf{ then }  $\bar{\mathbf{s}}^j_{t+1} \gets \bar{\mathbf{s}}^j_t - \mathbf{1}_{[l^{(k_j)}_t, r^{(k_j)}_t]}$.
    \end{ALC@g}
\ENDFOR
\end{algorithmic}
\end{algorithm}

\tw{
\subsubsection*{Performance Analysis.}
In this section, we establish the approximation guarantees of 
Algorithm~\ref{alg:assortment_general} for both the accept-or-reject and 
BAM-based scenarios under general arrivals.  
We begin with a lemma that lower bounds the expected revenue in a period when 
each assortment is offered with an attenuation probability.
}

\tw{
\begin{lemma}
\label{lemma:general 1 - 1/e}
In Algorithm~\ref{alg:assortment_general}, in any period \( t \in \mathcal{T} \),
and conditioned on the assortments \( \{ \mathcal{S}^{(k)} \}_{k \in [K_t]} \),
the expected revenue in period \( t \) is at least 
\(
(1 - 1/e) \sum_{k \in [K_t]} p^{(k)}_t \cdot R_t^{(k)}.
\)
\end{lemma}
}

\tw{
\proof{Proof of Lemma~\ref{lemma:general 1 - 1/e}.}
We apply the classical inequality of \citet{fleischer2011tight}:
}

\tw{
\begin{lemma}[{\sc Lemma 2.1 in \citealt{fleischer2011tight}}]
\label{lemma:folklore_inequality}
If \( f_1 \ge \cdots \ge f_M \ge 0 \), \( Y_i \ge 0 \), and 
\( \sum_{i=1}^M Y_i \le 1 \), then
\[
f_1 Y_1 + f_2 (1 - Y_1) Y_2 + \cdots + f_M \prod_{i=1}^{M-1} (1 - Y_i) Y_M 
\;\ge\; 
\left(1 - \left(1 - \frac{1}{M}\right)^M\right) \sum_{i=1}^M f_i Y_i.
\]
\end{lemma}
}

\tw{Conditioned on the assortments \( \{ \mathcal{S}^{(k)} \}_{k \in [K_t]} \), the expected revenue in
period \( t \) satisfies
\begin{align*}
    \sum_{i \in [K_t]} p^{(\phi_i)}_t \gamma_i R_t^{(\phi_i)}
    \;\ge\;
    \left(1 - \left(1 - \frac{1}{K_t}\right)^{K_t}\right)
    \sum_{k \in [K_t]} p^{(k)}_t R_t^{(k)}
    \;\ge\;
    \left(1 - \frac{1}{e}\right)
    \sum_{k \in [K_t]} p^{(k)}_t R_t^{(k)},
\end{align*}
where the first inequality applies Lemma~\ref{lemma:folklore_inequality} using the
fact that \( R_t^{(\phi_i)} \) is nonincreasing in \( i \) and the fact that $\sum_{i \in [K_t]} p^{(\phi_i)}_t = 1$, and the second inequality
uses \( 1 - (1 - 1/M)^M \ge 1 - 1/e \) for any \( M \ge 1 \).
\hfill \Halmos
}

\tw{Combining this lemma with the analysis of 
\( \mathbb{E}[R_t^{(k)}] \) (using 
the bound in~\eqref{eq:lower_bound_S} for the BAM-based scenario and  
the bound in~\eqref{eq:lower_bound_revenue_period} for the accept-or-reject 
scenario) yields the following approximation ratios (formal proofs appear in
Appendix~\ref{subsec:general_arrival_proof}):
}

\tw{
\begin{theorem}
\label{thm:BAM_general}
Algorithm~\ref{alg:assortment_general} achieves an approximation ratio of 
\minor{\( 0.271(1 - 1/e) \)} for the BAM-based scenario under general arrivals.
\end{theorem}
}

\tw{
\begin{theorem}
\label{thm:accept_or_reject_general}
For the BAM-based scenario under general arrivals with \( v^{(k)}_{t0} = 0 \),  
Algorithm~\ref{alg:assortment_general} achieves a  
\( (1 - 1/e)^2 \)-approximation ratio.  
By Lemma~\ref{lemma:special_case_general}, this implies that the accept-or-reject scenario under general arrivals also admits a  
\( (1 - 1/e)^2 \)-approximation policy.
\end{theorem}
}

\begin{remark}
\tw{In the online matching problem, the proposal-discarding framework of \cite{braverman2025new} (or the proposal-only algorithm of \citealt{braverman2022max}) can be extended from Bernoulli arrivals to general arrivals without any loss in the approximation ratio. The key reason is that these algorithms ultimately do not rely on discarding: \cite{braverman2025new} only introduce discarding to enforce probabilistic independence (or negative cylinder dependence) for analytical convenience under Bernoulli arrivals. \cite{braverman2022max} and \cite{braverman2025new} can show that even without discarding, the matched/unmatched statuses of offline nodes satisfy a form of negative dependence, though the probabilities of nodes being available can be larger than what is suggested by the LP.  
In contrast, in our setting, the availability of each resource is determined by its maximal sequence(s), making the evolution of its status substantially more intricate. As a result, we must use discarding to preserve the exact marginal probabilities suggested by the LP. 
Moreover, we are not aware of a suitable notion of negative dependence applicable to this more complex status structure.
Thus, we still keep probabilistic independence across resources.
Finally, to properly account for the randomness arising from multiple customer types, we introduce attenuation probabilities, which incur an additional performance loss.}
\end{remark}

\section{Conclusion}
\tw{In this paper, we extend the proposal-discarding framework to the network revenue management problem with consecutive stays and establish the first constant-factor approximation ratios for these settings. Several intriguing open questions remain.  
First, can the negative dependence techniques used in \cite{braverman2025new} be adapted to improve our guarantees under general arrivals? Addressing this would likely require resolving the distortion between LP-prescribed proposal probabilities and the actual marginal probabilities of maximal sequences.  
Second, in the BAM-based scenario, it would be valuable to conduct a more nuanced analysis of the revenue-maximizing assortment from the proposal set, and perhaps identify a tighter ratio.  
Third, it remains an open question whether we can extend the proposal-discarding framework to general choice models and obtain constant-factor approximation ratios in that broader setting. This may require understanding whether the fluid relaxation is tractable or admits a PTAS, together with suitable assortment-sampling techniques.  
Lastly, when restricted to \(L = 1\), it is natural to ask whether one can develop a philosopher inequality for online assortment optimization that surpasses the \(1/2\) barrier established in \cite{ma2021dynamic}.
}

\ACKNOWLEDGMENT{
The authors thank the Area Editor, Professor Ilan Lobel, the Associate Editor, and two anonymous reviewers for their constructive comments, which significantly improved the paper. We are particularly grateful to Referee 1 for suggesting a refinement of the performance guarantee in  Theorem~2 from $0.25$ to $0.271$, which also improves Theorem 3 accordingly. 
We also thank Referee 2 for suggesting a simplification of the proof of Theorem~\ref{thm:1 - 1/e} using the correlation gap.
The authors contributed equally and are listed alphabetically.
}

\bibliographystyle{informs2014}
\bibliography{refs}

\bigskip

\textbf{Ming Hu} is the University of Toronto Distinguished Professor of
Business Operations and Analytics and a professor of operations
management at the Rotman School of Management, University of
Toronto. His recent research interests include data-driven operations
and the interface of artificial intelligence and operations.

\textbf{Tongwen Wu} is a postdoctoral fellow at the Rotman School of Management, University of Toronto. He received his Ph.D. from the Department of Decision Analytics and Operations at City University of Hong Kong. His research interests include revenue management, online algorithms, approximation algorithms, and artificial intelligence.



\ECSwitch


\ECHead{\centering Online Appendix to ``Constant-Factor Algorithms for Revenue Management with Consecutive Stays''}
\begin{center}
    Ming Hu, Tongwen Wu
\end{center}

\bigskip

\section{Omitted Proofs in Section~\ref{sec:model}}
\label{sec:proof_special_case}
\proof{Proofs of Lemma~\ref{lemma:special_case} and~\ref{lemma:special_case_general}.}
Given an instance \(\mathcal{I}\) of the accept-or-reject scenario under Bernoulli (or general) arrivals, we construct a corresponding instance \(\mathcal{I}'\) of the BAM-based scenario by using the same parameters \( p_t, l_t, r_t \) and \(\{ w_{tj}\mid j \in \mathcal{M}\}\) (or \(\{ w^{(k)}_{tj}\mid j \in \mathcal{M}, k \in [K_t]\}\)). The choice model \(\{v_{tj}\mid j \in \mathcal{M}^+\}\) (or \(\{v^{(k)}_{tj}\mid j \in \mathcal{M}^+\}\)) for customer \(t\) is defined by setting \(v_{tj}=1\) (or $v^{(k)}_{tj} = 1$) for all \( j \in \mathcal{M}\) and \( v_{t0}=0 \) (or $v^{(k)}_{t0} = 0$). 
Because \( v_{t0}=0 \) (or $v^{(k)}_{t0} = 0$), the optimal policy for \(\mathcal{I}\) is also feasible for \(\mathcal{I}'\), as the customer's choice is deterministic when offered a single resource. Hence, \(\overline{V}(\mathcal{I}) \leq \overline{V}(\mathcal{I}')\). Furthermore, any \(\alpha\)-approximation policy for \(\mathcal{I}'\) can be transformed into an \(\alpha\)-approximation policy for \(\mathcal{I}\) by replacing each offered assortment \(\mathcal{S}\) with offering each resource in \(\mathcal{S}\) in equal chance. Thus, this policy is an $\alpha$-approximation policy for $\mathcal{I}$.
\hfill \Halmos

\section{Omitted Proofs in Section~\ref{sec:single_item}}
\label{subsec:proof_in_reject_or_accept}

\subsection{Omitted Proofs}
\proof{Proof of Lemma~\ref{lemma:upper}.}
Given an instance $\mathcal{I}$ of the accept-or-reject scenario under Bernoulli arrivals, let $\pi^* = \arg \max_{ \pi } V^{\pi}(\mathcal{I})$ be the optimal policy that maximizes the total expected revenue. Under the optimal policy $\pi^*$, let $\mathbf{s}^j_t$ denote the available status of resource $j \in \mathcal{M}$ at the beginning of period $t$, $\bar{x}^j_t([a,b])$ denote the probability that $[a,b]\sim \mathbf{s}^j_t$, and 
\[ \bar{y}^j_t([a,b]) = \begin{cases}
    \Pr \{ [a,b] \sim \mathbf{s}^j_t \text{ and } [l_t,r_t] \text{ is allocated to request } t \}, & \text{ if } [l_t,r_t] \subseteq [a,b], \\
    0, & \text{ otherwise. }
\end{cases}\] 
Next, we show $(\bar{\mathbf{x}}, \bar{\mathbf{y}})$ is a feasible solution to \eqref{eq:lp_single}. 
Since request $t$'s arrival is independent of the initial resource status at the period $t$, we have $\bar{y}^j_t([a,b]) \le \bar{x}^j_t([a,b]) \cdot p_t$, satisfying Constraint \eqref{eq:allocate_single}. 
By the definition of $(\bar{\mathbf{x}}, \bar{\mathbf{y}})$, they also satisfy Constraints \eqref{eq:no_allocate} and \eqref{eq:boundary_single}. 
Constraint \eqref{eq:trans_single} is naturally satisfied because it accounts for the probability mass flow across periods. Constraint \eqref{eq:across_single} is satisfied since the customer arrival probability is $p_t$. Thus, the feasibility is established. Finally, since $V^{\pi^*}(\mathcal{I}) = \sum_{j \in \mathcal{M}} \sum_{t \in \mathcal{T}} \sum_{1 \le a \le b \le N} w_{tj} \bar{y}^j_t([a,b])$, we conclude that $\overline{V}(\mathcal{I}) = V^{\pi^*}(\mathcal{I}) \le \texttt{LP}(\mathcal{I})$.  
\hfill \Halmos

\proof{Proof of Proposition \ref{prop:virtual_proerty}.}
We prove the result by induction on $t \in \mathcal{T}$.

\textbf{Base case ($t = 1$):}
By initialization, $\bar{\mathbf{s}}^j_1 = \mathbf{s}^j_1 = \mathbf{1}_{[1,N]}$ for all $j \in \mathcal{M}$, and thus the only maximal sequence is $[1,N]$. Therefore, we have
\(
\Pr\{ [a,b] \sim \bar{\mathbf{s}}^j_1 \} = \mathbbm{1}\{ [a,b] = [1,N] \} = x_{1j}([a,b]).
\)
Moreover, since all virtual statuses are deterministic at initialization, independence across resources trivially holds.

\textbf{Inductive step:} Assume the property holds for all periods up to \(t'\). We show it holds for \(t'+1\).

(i) Lower Bound: 
If resource $j^* \in \mathcal{P}$ is selected for allocation, its virtual status is updated as
\(
\bar{\mathbf{s}}^{j^*}_{t'+1} = \bar{\mathbf{s}}^{j^*}_{t'} - \mathbf{1}_{[l_{t'}, r_{t'}]},
\)
which mirrors the actual allocation. Thus, $\bar{\mathbf{s}}^{j^*}_{t'+1} \le \mathbf{s}^{j^*}_{t'+1}$. For any $j \in \mathcal{M} \setminus \{j^*\}$, the virtual status is either unchanged or updated by subtracting $\mathbf{1}_{[l_{t'}, r_{t'}]}$, yielding
\(
\bar{\mathbf{s}}^j_{t'+1} \le \bar{\mathbf{s}}^j_{t'}.
\)
By the induction hypothesis, $\bar{\mathbf{s}}^j_{t'} \le \mathbf{s}^j_{t'} = \mathbf{s}^j_{t'+1}$, and therefore
$
\bar{\mathbf{s}}^j_{t'+1} \le \mathbf{s}^j_{t'+1}.
$

(ii) Marginal Probability: 
We show that
$
\Pr\{ [a,b] \sim \bar{\mathbf{s}}_{t'+1,j} \} = x_{t'+1,j}([a,b])
$
holds for all $1 \le a \le b \le N$. This probability consists of two contributions: the probability that $[a,b]$ was a maximal sequence in period $t'$ and remained unchanged, and the probability that $[a,b]$ was newly formed by a split due to an allocation.
The former is given by:
$$
\Pr\{ [a,b] \sim \bar{\mathbf{s}}^j_{t'}, \text{ and unchanged} \} = x_{t'j}([a,b]) - y_{t'j}([a,b]),
$$
since $y_{t'j}([a,b])$ corresponds to the probability that the allocation occurs.
For the latter, $[a,b] \sim \bar{\mathbf{s}}^j_{t'+1}$ may arise from a breakup of a longer maximal sequence $[a',b']$, where $[a,b] = [a', l_{t'} - 1]$ or $[a,b] = [r_{t'} + 1, b']$.
These contributions are:
$
\sum_{r_{t'} \le b' \le N} \mathbbm{1}\{ b = l_{t'} - 1 \} y_{t'j}([a, b']) + \sum_{1 \le a' \le l_{t'}} \mathbbm{1}\{ r_{t'} + 1 = a \} y_{t'j}([a', b]).
$
Adding both contributions yields
$
\Pr\{ [a,b] \sim \bar{\mathbf{s}}^j_{t'+1} \} = x_{t'+1,j}([a,b])$, which follows directly from Equality \eqref{eq:trans_single}.

(iii) Independence: 
Proposals and updates in the algorithm are executed independently across resources. By the induction hypothesis, the collections $\{ \bar{\mathbf{s}}^j_t \}_{1 \le t \le t'}$ are independent across $j \in \mathcal{M}$, and the independence is preserved by the update procedure. Thus,
$
\{ \bar{\mathbf{s}}^j_t \}_{1 \le t \le t'+1}$ are independent across $j \in \mathcal{M}.
$
\hfill \Halmos

\subsection{NP-Hardness}
\label{sec:hardness}

\color{black}

\proof{Proof of Proposition~\ref{prop:hardness_accept_reject}.}

We provide a gap-preserving reduction from the MAX-3SAT problem to an offline instance of the accept-or-reject scenario under Bernoulli arrivals, where all arrival probabilities \( p_t = 1 \). The definition and hardness result of MAX-3SAT are as follows:

\begin{definition}[MAX-3SAT]
\label{dfn:3-sat}
Given \( n \) Boolean variables \( (x_i)_{i=1}^n \) and \( m \) clauses \( (C_k)_{k=1}^m \) in 3-CNF form, where each clause contains at most three literals (i.e., variables or their negations), the MAX-3SAT problem seeks an assignment that satisfies the maximum number of clauses.
\end{definition}

\begin{lemma}[\cite{haastad2001some}]
For any \( \epsilon > 0 \), it is NP-hard to distinguish whether a 3-CNF formula is fully satisfiable or whether at most a \( \tfrac{7}{8} + \epsilon \) fraction of clauses can be satisfied.
\end{lemma}

Let us construct an instance of the accept-or-reject scenario from a given 3-CNF formula with \( n \) variables and \( m \) clauses. Let \( c_i \) denote the number of occurrences of \( x_i \) or \( \neg x_i \) across all clauses. We set the number of resources to \( M = 2n \) and the number of slots to \( N = m \). Since all \( p_t = 1 \), we may ignore the arrival order and omit the time index \( t \) for notational simplicity.

We define two types of requests:
\begin{itemize}
    \item[(i)]Variable requests:
For each variable \( x_i \), we introduce a request:
\[
\tilde{\theta}_i = (\tilde{p}_i = 1, \tilde{l}_i = 1, \tilde{r}_i = N,
\{ \tilde{w}_{i, 2i-1} = \tilde{w}_{i, 2i} = c_i/2, \tilde{w}_{ij} = 0 \text{ for } j \ne 2i-1, 2i \}).
\]
That is, \( \tilde{\theta}_i \) can be allocated to either resource \( 2i-1 \) or \( 2i \), with revenue \( c_i/2 \), consuming all the slots of the assigned resource.

\item[(ii)] Clause requests:
For each clause \( C_k \), we define a request:
\[
\hat{\theta}_k = (\hat{p}_k = 1, \hat{l}_k = \hat{r}_k = k,
\{ \hat{w}_{kj} | j \in \mathcal{M}\} ), \text{ where }
\]
\[
\hat{w}_{kj} = \begin{cases}
1, & \text{if } j = 2i \text{ and } x_i \in C_k, \\
1, & \text{if } j = 2i-1 \text{ and } \neg x_i \in C_k, \\
0, & \text{otherwise}.
\end{cases}
\]
In other words, clause requests can be assigned only to the slot corresponding to the clause index, and only to resources representing the literals appearing in the clause.
\end{itemize}

We now argue the two sides of the gap:
\paragraph{YES case (fully satisfiable):}
Suppose all clauses can be satisfied. Then, for each \( x_i \), we allocate \( \tilde{\theta}_i \) to resource \( 2i \) if \( x_i \) is false, or to resource \( 2i-1 \) if \( x_i \) is true. For each satisfied clause \( C_k \), we assign \( \hat{\theta}_k \) to an available slot of the corresponding resource (either \( 2i \) or \( 2i-1 \)) that represents a satisfied literal. Since every clause is satisfied, this yields a total revenue of:
\[
\sum_{i=1}^n \frac{c_i}{2} + m.
\]

\paragraph{NO case (at most \( \tfrac{7}{8} + \epsilon \) of clauses satisfiable):}
First, we must still assign each \( \tilde{\theta}_i \) to either \( 2i-1 \) or \( 2i \) to capture the full \( c_i/2 \) revenue, since the total potential clause revenue from \( \hat{\theta}_k \) involving \( x_i \) is at most \( c_i \). Once the variable requests are placed, a variable assignment is decided. Given this assignment, we can satisfy at most \(( \tfrac{7}{8} + \epsilon)m \) clauses. So, the maximum total revenue is:
\[
\sum_{i=1}^n \frac{c_i}{2} + \left(\tfrac{7}{8} + \epsilon\right)m.
\]

Finally, since each clause contains at most three literals, \( \sum_{i=1}^n c_i \le 3m \). Thus,
\[
\frac{\sum_i c_i/2 + (7/8 + \epsilon)m}{\sum_i c_i/2 + m}
\le \frac{3m/2 + (7/8 + \epsilon)m}{3m/2 + m} = \frac{(19/8 + \epsilon)m}{(5/2)m} = \frac{19}{20} + O(\epsilon).
\]

Hence, a \( \left( \tfrac{19}{20} + O(\epsilon) \right) \)-approximation algorithm for the accept-or-reject scenario (with \( p_t = 1 \)) would imply an algorithm that distinguishes between fully satisfiable and partially satisfiable 3-CNF formulas, which is NP-hard.
\hfill \Halmos

\subsection{Integrality Gap}
\label{sec:integrality_gap}

\proof{Proof of Proposition~\ref{prop:integrality_gap_accept_reject}.}

\citet{papadimitriou2024online} show that the integrality gap of the linear relaxation when  \(L = 1\) is at most \(1 - \frac{1}{2e} \approx 0.816\). 
We extend their construction to cases where \(L = 2\), thereby deriving a slightly smaller upper bound on the integrality gap.

\paragraph{Construction of the instance.}
Consider an instance \(\mathcal{I}_q\) with \(M = 2\) resources and \(N = 2\) slots for an integer \(q \ge 1\).  
Let \(q' = \lfloor (1 - \ln 2) q \rfloor.\)  
The revenue from fulfilling a request equals the number of slots it occupies. 
We define four types of requests:

\begin{itemize}
    \item[(i)] {1-Requests:}  
    The first \(q'\) requests are identical, each defined as
    \begin{align*}
    \theta_t = (p_t = 1/q,\, l_t = r_t = 1,\, \{ w_{t1} = w_{t2} = 1 \}), \quad \forall\, 1 \le t \le q'.
    \end{align*}
    Each 1-request demands only slot 1 and can be fulfilled by either resource, arriving with probability \(1/q\).
    \item[(ii)] {2-Requests:}  
    The next \(q'\) requests are identical:
    \[
    \theta_t = (p_t = 1/q,\, l_t = r_t = 2,\, \{ w_{t1} = w_{t2} = 1 \}), \quad \forall\, q' + 1 \le t \le 2q'.
    \]
    Each 2-request demands only slot 2 and can be fulfilled by either resource, also arriving with probability \(1/q\).

    \item[(iii)] {(1,2)-Requests:}  
    The following \(q - q'\) requests are of the same type:
    \[
    \theta_t = (p_t = 1/q,\, l_t = 1,\, r_t = 2,\, \{ w_{t1} = w_{t2} = 2 \}), \quad \forall\, 2q' + 1 \le t \le q' + q.
    \]
    Each of these requests demands both slots and generates revenue 2, also arriving with probability $1/q$.

    \item[(iv)] {Final Request:}  
    The last request is deterministic:
    \[
    \theta_{q' + q + 1} = (p_{q' + q + 1} = 1,\, l_{q' + q + 1} = 1,\, r_{q' + q + 1} = 2,\, \{ w_{q'+q+1, 1} = w_{q'+q+1, 2} = 2 \}).
    \]
\end{itemize}

\paragraph{Optimal value of the dynamic program \(\overline{V}(\mathcal{I}_q)\).}
It is clear that the optimal policy allocates the first \(q' + q\) stochastic requests to one resource and reserves the other resource for the last deterministic request. 
Thus, we need only consider the optimal policy for a single resource with the first \(q' + q\) requests.

When a 2-request arrives and no other request has yet been accepted, accepting it yields a revenue of 1, while rejecting it preserves the chance to accept a (1,2)-request later.  
The expected future revenue from rejecting is at least \(2 \cdot (1 - (1 - 1/q)^{q - q'})\), since a (1,2)-request can be accepted once one arrives.  
Because \(2 \cdot (1 - (1 - 1/q)^{q - q'}) \ge 2(1 - (1-1/q)^{q\ln 2 }) \ge 1\), it is optimal to reject 2-requests as long as no 1-request has been accepted.

For the first 1-request to arrive, the expected revenue from accepting it is \(1 + (1 - (1 - 1/q)^{q'})\), since after accepting, one can still accept any arriving 2-request.  
The expected revenue from rejecting it is \(2 \cdot (1 - (1 - 1/q)^{q - q'})\), as noted earlier.  
For large \(q\), \(2 \cdot (1 - (1 - 1/q)^{q - q'}) \to 1\), while \(1 + (1 - (1 - 1/q)^{q'}) > 1\), so accepting the first 1-request to arrive is optimal.

Hence, when \(q\) is sufficiently large, the optimal policy is:
\begin{itemize}
    \item If at least one 1-request arrives,  accept the first 1-request to arrive, and any subsequent 2-request (if one arrives);
    \item If no 1-request arrives, accept the first (1,2)-request (if any).
\end{itemize}

Thus, for large \(q\),
\begin{align*}
\overline{V}(\mathcal{I}_q)
&= 2 + \Big(1 - (1 - 1/q)^{q'}\Big)\Big(1 + (1 - (1 - 1/q)^{q'})\Big)
   + (1 - 1/q)^{q'} \Big(1 - (1 - 1/q)^{q - q'}\Big) \cdot 2 \\
&\to 2 + \left(1 - \frac{2}{e}\right)\left(1 + 1 - \frac{2}{e}\right) + \frac{2}{e}.
\end{align*}

\paragraph{Optimal value of the linear relaxation \(\emph{\texttt{LP}}(\mathcal{I}_q)\).}
We construct a feasible fractional solution to provide a lower bound on \(\texttt{LP}(\mathcal{I}_q)\).  
Consider a policy for a single resource that accepts:
1-requests, (1,2)-requests, and the final deterministic request with ex-ante probabilities \(1/(2q)\), \(1/(2q)\), and \(1/2\), respectively.  
This is feasible since the probability that no request is accepted (conditioned on an arrival) is at least \(1/2\).
For each 2-request, the policy accepts it if a 1-request has already been accepted and no 2-request has been accepted yet.  
We set the variables \(\mathbf{x}\) and \(\mathbf{y}\) for each resource \(j \in \{1,2\}\) according to this policy.
To verify the feasibility, Constraint \eqref{eq:across_single} is satisfied since, for each type, the ex-ante allocation probability of each request to each resource is at most half of the arrival probability. The other constraints are satisfied by our construction because they arise from the single-resource policy. Hence,
\begin{align*}
\texttt{LP}(\mathcal{I}_q)
&\ge 2 \cdot \Bigg[
    \frac{1}{2q} \cdot q' \cdot \Big(1 + (1 - (1 - \tfrac{1}{q})^{q'})\Big)
    + \frac{1}{2q} (q - q') \cdot 2
    + \frac{1}{2} \cdot 2
\Bigg] \\
&\to 2 \cdot \Bigg[
    \frac{1 - \ln 2}{2} \cdot \Big(2 - \frac{2}{e}\Big)
    + \ln 2 + 1
\Bigg].
\end{align*}

\paragraph{Conclusion.}
Therefore, the integrality gap satisfies
\[
\inf_{\mathcal{I}} \Big\{ \frac{\overline{V}(\mathcal{I})}{\texttt{LP}(\mathcal{I})} \Big\} \le 0.8134.
\]
\hfill \Halmos

\color{black}

\section{Omitted Details in Section~\ref{sec:assortment}}
\label{sec:appendix_assortment}

\subsection{Omitted Proofs}
\label{subsec:proofs_assortment}
\proof{Proof of Lemma~\ref{lemma:upper_assortment}.}
The proof is similar to the proof of Lemma \ref{lemma:upper} except additionally utilizing the property of the basic attraction model. Let $\pi^* = \arg \max_{ \pi  } V^{\pi}(\mathcal{I})$ be the optimal policy given the instance $\mathcal{I}$. Under the optimal policy $\pi^*$, let $\mathbf{s}^j_t$ denote the available status of resource $j \in \mathcal{M}$ at the beginning of period $t$, $\bar{x}^j_t([a,b])$ denote the probability that $[a,b]\sim \mathbf{s}^j_t$, 
\[ \bar{y}_{tj}([a,b]) = \begin{cases}
    \Pr \{ [a,b] \sim \mathbf{s}^j_t \text{ and customer } t \text{ chooses resource } j  \}, & \text{ if } [l_t,r_t] \subseteq [a,b], \\
    0, & \text{ otherwise. }
\end{cases}\] 
and
\[ \bar{y}^{0}_{tj}([a,b]) = \begin{cases}
    \Pr \{ [a,b] \sim \mathbf{s}^j_t \text{, resource } j \text{ is offered to customer } t \text{, and she chooses } 0 \}, & \text{ if } [l_t,r_t] \subseteq [a,b], \\
    0, & \text{ otherwise. }
\end{cases}\] 
Let $\bar{y}^0_t$ be the probability that customer $t$ arrives and chooses the outside option.
Next, we show $(\bar{\mathbf{x}}, \bar{\mathbf{y}})$ is a feasible solution to \eqref{eq:sblp}. 
Constraint \eqref{eq:scale} is satisfied since each time interval \([a,b]\sim \bar{\mathbf{s}}^j_t\) and resource \( j \) is offered to customer \( t \), the choice probabilities of resource \( j \) and the outside option follow the fixed ratio \(\frac{v_{tj}}{v_{t0}}\).  
Since $\bar{y}^{0}_{tj}([a,b]) + \bar{y}_{tj}([a,b])$ is no more than the probability that $[a,b] \sim \bar{\mathbf{s}}^j_t$ and the resource $j$ is offered to customer $t$, Constraint \eqref{eq:allocate_assortment} holds. Constraints \eqref{eq:no_allocate_assortment}, \eqref{eq:balance} and \eqref{eq:boundary_assortment} naturally hold for any policy. Constraints \eqref{eq:no_purchase} and \eqref{eq:capacity} hold by the definitions of $\bar{y}_{tj}(\cdot), \bar{y}^{0}_{tj}(\cdot) $ and $\bar{y}^0_t$.
\hfill \Halmos

\proof{Proof of Proposition \ref{prop:virtual_property_assortment}.}
We prove the result by induction on $t$. 

\textbf{Base case ($t = 1$):} 
By initialization, the real and virtual resource statuses are identical: $\bar{\mathbf{s}}^j_1 = \mathbf{s}^j_1 = \mathbf{1}_{[1,N]}$ for all $j \in \mathcal{M}$. The only maximal sequence is $[1, N]$, hence $\Pr\{ [a, b] \sim \bar{\mathbf{s}}^j_1 \} = x_{1j}([a, b]) = \mathbbm{1}\{[a, b] = [1, N]\}$. As the statuses are deterministic, independence across resources is trivially satisfied. Moreover, since all resources are initially available, the candidate assortment is feasible.

\textbf{Inductive step}:
Assume Property~\ref{property:virtual} and the feasibility of candidate assortments hold for all periods up to $t'$. We show they also hold for period $t'+1$.
Let $j^*$ denote the realized selection in period $t'$ or the outside option if customer $t'$ does not arrive. We verify the following:

(i) Lower Bound: 
For any $j \neq j^*$, we have
$
\bar{\mathbf{s}}^j_{t'+1} \le \bar{\mathbf{s}}^j_{t'} \le \mathbf{s}^j_{t'} = \mathbf{s}^j_{t'+1}.
$
For resource $j^*$ that the customer selects, by the Inclusion property of Proposition~\ref{prop:random}, $j^* \in \mathcal{Q}$, and thus:
$
\bar{\mathbf{s}}^{j^*}_{t'+1} = \bar{\mathbf{s}}^{j^*}_{t'} - \mathbf{1}_{[l_{t'}, r_{t'}]} \le \mathbf{s}^{j^*}_{t'} - \mathbf{1}_{[l_{t'}, r_{t'}]} = \mathbf{s}^{j^*}_{t'+1}.
$

(ii) Feasibility of $\mathcal{S}$ in period $t'+1$: 
In period $t'+1$, the candidate assortment $\mathcal{S}$ is formed as a subset of $\mathcal{P}$, the set of resources that submit proposals. If $j \in \mathcal{P}$, then there exists a maximal sequence $[a,b] \sim \bar{\mathbf{s}}^j_{t'+1}$ such that $[l_{t'+1}, r_{t'+1}] \subseteq [a,b]$. Since $\bar{\mathbf{s}}^j_{t'+1} \le \mathbf{s}^j_{t'+1}$, the feasibility follows.

(iii) Independence: 
Given the proposal set $\mathcal{P}$ and the candidate assortment $\mathcal{S}$, the update set $\mathcal{Q}$ is generated as $\textsc{Random}(\mathbf{q}, \mathbf{q}', j^*)$, where $j^* \sim P$, and $P$ is defined by $\mathbf{q}'$:
$
\Pr\{ j^* = j \} = q'_j, \ \forall j \in \mathcal{M}.
$
To verify that $\mathbf{q}$ and $\mathbf{q}'$ satisfy the condition \eqref{eq:regular_condtion}, observe that for all $j \in \mathcal{M}$,
$$
\frac{q'_j}{1 - \sum_{j'=j+1}^M q'_{j'}} 
= \frac{\mathbbm{1}\{ j \in \mathcal{S} \} p_t v_{tj}}{v_{t0} + \sum_{j' \in \mathcal{S}} v_{tj'} - \sum_{j'=j+1}^M \mathbbm{1}\{ j' \in \mathcal{S} \} p_t v_{tj'}} 
\le \mathbbm{1}\{ j \in \mathcal{S} \} \cdot \frac{p_t v_{tj}}{v_{t0} + v_{tj}} 
\le q_j.
$$
Above, the first equality follows from \eqref{eq:q'}. The first inequality holds trivially if $j \notin \mathcal{S}$; and if $j \in \mathcal{S}$, it holds because the denominator on the right-hand side is no more than that of the left-hand side. The last inequality uses the fact that $\mathcal{S} \subseteq \mathcal{P}$.
Hence, by Proposition~\ref{prop:random}, the output set $\mathcal{Q}$ includes each resource independently with probability $q_j$, conditioned on $\mathcal{P}$ and $\mathcal{S}$.
Moreover, since the inclusion of resource $j$ in $\mathcal{P}$, as well as the value of $q_j$, are all independent of other resources, the independence of the virtual resource statuses is preserved.

(iv) Marginal Probability: 
Conditioned on $[a, b] \sim \bar{\mathbf{s}}^j_{t'}$ with $[l_{t'}, r_{t'}] \subseteq [a, b]$, the probability that the virtual status of resource $j$ is updated is:
$$
\frac{y^{0}_{t'j}([a, b]) + y_{t'j}([a, b])}{x_{t'j}([a, b]) \cdot p_{t'}} \cdot p_{t'} \cdot \frac{v_{t'j}}{v_{t'0} + v_{t'j}} = \frac{y_{t'j}([a, b])}{x_{t'j}([a, b])},
$$
where the equality follows from Constraint~\eqref{eq:scale}. Together with Equation~\eqref{eq:balance}, this confirms the correct marginal distribution in period $t'+1$.
\hfill \Halmos

\subsection{Details of \textsc{Random}($\mathbf{q}, \mathbf{q}',  \tilde{j}$)}
\label{subsec:random}
The implementation of $\textsc{Random}(\mathbf{q}, \mathbf{q}', \tilde{j})$ is provided in Algorithm~\ref{alg:random}. The procedure processes the resources in $\mathcal{M}$ sequentially in the reverse order, from $M$ down to $1$. At each step $j$, the algorithm decides whether to include resource $j$ based on $\tilde{j}$ and a randomized decision rule governed by $\mathbf{q}$ and $\mathbf{q}'$. We next establish Proposition~\ref{prop:random}.

\begin{algorithm}
\caption{\textsc{Random}($\mathbf{q}, \mathbf{q}',  \tilde{j}$)}
\label{alg:random}
\begin{algorithmic}[1]
    \REQUIRE Probability vectors $\mathbf{q}, \mathbf{q}' \in [0,1]^M$, and a choice $\tilde{j}$.
    \ENSURE $\mathcal{Q}$
    \STATE Initialize $\mathcal{Q} \leftarrow \emptyset$.
    \FOR{$j=M$ \textbf{downto} $1$}
    \STATE Let $z_j \gets \frac{q'_j}{1 - \sum_{j'=j+1}^M q'_{j'}}$.
    \SWITCH {$j$}
    \CASE {$ j > \tilde{j}$}
    \STATE Add $j$ into $\mathcal{Q}$ with probability $ \max \{0, \min \{ 1,(q_j - z_j)/(1-z_j) \} \}$ independently.
    \ENDCASE
    \CASE {$j = \tilde{j}$}
    \STATE Add $j$ into $\mathcal{Q}$.
    \ENDCASE
    \CASE {$j < \tilde{j}$}
    \STATE Add $j$ into $\mathcal{Q}$ with probability $ q_j$ independently.
    \ENDCASE
    \ENDSWITCH
    \ENDFOR
\end{algorithmic}
\end{algorithm}

\proof{Proof of Proposition~\ref{prop:random}.}

The time complexity of Algorithm~\ref{alg:random} is clearly \( O(M) \), as it consists of a single loop running \( M \) iterations. If \( \tilde{j} = j \in \mathcal{M} \), then resource \( j \) is included in \( \mathcal{Q} \) at iteration \( j \) (by line 8 of Algorithm~\ref{alg:random}). 

Next, we focus on proving the independence. Suppose inputs \(\mathbf{q}, \mathbf{q}' \in [0,1]^M\) are fixed so that they satisfy inequality \eqref{eq:regular_condtion}, and the random input \(\tilde{j}\) follows the distribution \(P\) defined in \eqref{eq:distribution}.  \(\mathcal{Q}\) denotes the random output set. 
To show that each resource \(j \in \mathcal{M}\) is included independently in \(\mathcal{Q}\) with probability \(q_j\), the central idea involves coupling the distribution of \(\tilde{j}\) with a backward indexed Markov chain process \(\{Z_j\}_{j=M,\dots,1}\), where the realization of \(Z_j\) is determined directly by \(\tilde{j}\) via \(Z_j = \mathbbm{1}[\tilde{j} \ge j]\). Based on the distribution of \(\tilde{j}\) defined in \eqref{eq:distribution}, the transition probabilities of this Markov chain can be explicitly computed as:
\begin{align}
\label{eq:markov_prob}
\begin{split}
& z_j \coloneqq \frac{q'_j}{1 - \sum_{j' = j+1}^M q'_{j'}} \quad \forall\, 1 \le j \le M, \\
& \Pr\{Z_M = 1\} = z_M, \quad \Pr\{Z_M = 0\} = 1 - z_M, \\
& \Pr\{Z_j = 1 \mid Z_{j+1} = 0\} = z_j, \quad \Pr\{Z_j = 0 \mid Z_{j+1} = 0\} = 1 - z_j, \quad \forall\, 1 \le j \le M-1, \\
& \Pr\{Z_j = a \mid Z_{j+1} = 1\} = \mathbbm{1}\{a = 1 \}, \quad \forall\, 1 \le j \le M-1,\, a \in \{0,1\}.
\end{split}
\end{align}
Indeed, we can directly infer the value of \(\tilde{j}\) from the realization of the Markov chain \(\{Z_j\}_{j=1,\dots,M}\). Thus, our algorithm can be thought of as coupling \(\mathcal{Q}\) with the Markov chain \(\{Z_j\}_{j=1,\dots,M}\). Specifically, given fixed inputs \((\mathbf{q}, \mathbf{q}')\), we sequentially realize the Markov chain from \(Z_M\) to \(Z_1\).
Alongside this realization, our algorithm determines sequentially, for each resource \(j \in \mathcal{M}\), whether it should be included in \(\mathcal{Q}\). Suppose we are currently at resource \(j \in \mathcal{M}\):
\begin{itemize}
    \item If \(j < M\) and \(Z_{j+1} = 1\), it implies \(\tilde{j} > j\) and thus \(Z_j = 1\); in this case, we include resource \(j\) in \(\mathcal{Q}\) with probability \(q_j\) (corresponding to line 10 in Algorithm~\ref{alg:random}).
    \item  Otherwise, the conditional probability that \(Z_j = 1\) is exactly \(z_j\) (as given by \eqref{eq:markov_prob}). We then couple the inclusion of \(j\) in \(\mathcal{Q}\) with \(Z_j\): if \(Z_j = 1\), we include \(j\) in \(\mathcal{Q}\) (corresponding to line 8 in Algorithm~\ref{alg:random}); if \(Z_j = 0\), we include resource \(j\) in \(\mathcal{Q}\) with probability \((q_j - z_j)/(1 - z_j)\) (corresponding to line 6 in Algorithm~\ref{alg:random}). Note that \((q_j - z_j)/(1 - z_j)\) lies within the interval \([0,1]\) since $z_j \le q_j \le 1$ if \(\mathbf{q}\) and \(\mathbf{q}'\) satisfy \eqref{eq:regular_condtion} and $\mathbf{q} \in [0,1]^M$. It is straightforward to verify that, in this case, the probability of including resource \(j\) is also \(q_j\). 
\end{itemize}

From the above explanation, we know that if vectors \( \mathbf{q}, \mathbf{q}' \in [0,1]^M \) satisfy condition \eqref{eq:regular_condtion} and the random input \( \tilde{j} \) follows distribution \( P \) defined in \eqref{eq:distribution}, the conditional probability of including resource \( j \) into \( \mathcal{Q} \) at each step is:
\begin{align*}
\Pr\left[ j \in \mathcal{Q} \,\middle|\, \{ Z_{j'} \}_{j'=j+1}^M,\, \mathcal{Q} \cap \{j+1,\dots,M\} \right] = q_j,
\end{align*}
which implies that $\Pr[ j \in \mathcal{Q} | \mathcal{Q} \cap \{j+1,\dots,M\} ] = q_j$.
Therefore, for any $\mathcal{X} \subseteq \mathcal{M}$,
\begin{align*}
           & {\Pr}_{\tilde{j} \sim P }\Big[  \mathcal{Q} 
            = \mathcal{X} \Big| \mathbf{q}, \mathbf{q}' \Big] \\
           &\  = \prod_{j=1}^M \Big( \mathbbm{1}[j \in \mathcal{X}] \Pr[j \in \mathcal{Q} | \mathcal{Q} \cap \{j+1,\dots,M\} = \mathcal{X} \cap \{j+1,\dots,M \} ] \\
           & \quad + \mathbbm{1}[j \notin \mathcal{X}] \Pr[j \notin \mathcal{Q} | \mathcal{Q} \cap \{j+1,\dots,M\} = \mathcal{X} \cap \{j+1,\dots,M \} ]  \Big) \\
           & = \prod_{j=1}^M \Big( \mathbbm{1}[j \in \mathcal{X}] q_j + \mathbbm{1}[j \notin \mathcal{X}] (1-q_j) \Big) \\
           & = \prod_{j \in \mathcal{X}} q_j \prod_{j \in \mathcal{M}\backslash \mathcal{X}} (1-q_j).  
\end{align*}
\hfill \Halmos

\color{black}

\subsection{Proof of the Technical Lemma~\ref{lemma:technical}}
\label{subsec:proof_technical_lemma}
\proof{Proof of Lemma~\ref{lemma:technical}.}

Fix \minor{$\rho>0$, $\lambda=1+1/\rho$, and} $L, S \ge 0$ with $L + S \le 1$. For these values, consider minimizing
\[
Z \triangleq \max \left\{ \minor{\frac{X}{(1+\rho)S + 1}},\ \minor{\frac{(1 - e^{-\lambda L})Y}{\lambda L}} \right\}
\quad \text{subject to } X, Y \ge 0, \ X + Y = 1.
\]
At a minimum, the two terms inside the maximum must be equal; otherwise, shifting probability mass from the larger to the smaller term would reduce the maximum. Hence,
\[
\minor{\frac{X}{(1+\rho)S + 1} = \frac{(1 - e^{-\lambda L})Y}{\lambda L} = Z.}
\]
Together with $X + Y = 1$, we obtain
\[
X = \minor{Z((1+\rho)S + 1)}, \qquad Y = \minor{Z \frac{\lambda L}{1 - e^{-\lambda L}}}, \qquad
\Rightarrow \quad
Z(L, S) = \minor{\frac{1}{1 + (1+\rho)S + \frac{\lambda L}{1 - e^{-\lambda L}}}}.
\]
Thus, for any feasible $X, Y$, 
\[
\max \left\{ \minor{\frac{X}{(1+\rho)S + 1}},\ \minor{\frac{(1 - e^{-\lambda L})Y}{\lambda L}} \right\} \ge Z(L, S).
\]
\minor{It remains to upper bound the denominator. Since $S \le 1-L$,}
\[
\minor{1 + (1+\rho)S + \frac{\lambda L}{1-e^{-\lambda L}}
\le 1 + (1+\rho)(1-L) + \frac{\lambda L}{1-e^{-\lambda L}}
\triangleq f_\rho(L).}
\]
\minor{The function $g(x)=x/(1-e^{-x})$ is convex on $x\ge 0$ because}
\[
\minor{g''(x)=\frac{e^{-x}\left(x-2+(x+2)e^{-x}\right)}{(1-e^{-x})^3}\ge 0,}
\]
\minor{where the inequality follows from $x-2+(x+2)e^{-x}\ge 0$ for $x\ge0$. Hence, $f_\rho(L)$ is convex in $L$ on $[0,1]$, and it is bounded above by the larger of its endpoint values. Taking the limit at $L=0$ gives}
\[
\minor{\lim_{L\downarrow0} f_\rho(L)=3+\rho,\qquad f_\rho(1)=1+\frac{\lambda}{1-e^{-\lambda}}.}
\]
\minor{Therefore,}
\[
\minor{1 + (1+\rho)S + \frac{\lambda L}{1-e^{-\lambda L}}
\le \max\left\{3+\rho,\ 1+\frac{\lambda}{1-e^{-\lambda}}\right\},}
\]
\minor{which proves the desired lower bound on $Z(L,S)$ and completes the proof.}
\hfill \Halmos

\subsection{Integrality Gap}
\proof{Proof of Proposition~\ref{prop:integrality_gap_BAM}.}
Let \(m\) range over positive integers satisfying \(m \equiv 4 \pmod 6\), and let
\(
q = \frac{2(m-1)}{3}.
\)
We construct an instance \(\mathcal{I}_m\) and evaluate \(\overline{V}(\mathcal{I}_m)\) and \(\texttt{SBLP}(\mathcal{I}_m)\), respectively.

\paragraph{Construction of \(\mathcal{I}_m\).}
Consider a setting with one slot (\(N = 1\)) and \(m\) resources (\(M = m\)).
Let the total number of requests be \(T = q+1\).
The first \(q\) requests are of the accept-or-reject type:
\[
\theta_t = \left(p_t = 1,\, l_t = r_t = 1,\, \left\{w_{tj} = \frac{1}{2m}\right\}_{j \in \mathcal{M}},\, \left\{v_{t0}=0,\, v_{tj}=1\right\}_{j \in \mathcal{M}}\right), \quad 1 \le t \le q.
\]
Each such request arrives with probability one, can be served using any available resource, and yields revenue \(1/(2m)\) if fulfilled.
The final request is of BAM-type:
\[
\theta_{t'} = \left(p_{t'} = 1,\, l_{t'} = r_{t'} = 1,\, \{w_{t'j}=1\}_{j \in \mathcal{M}},\, \left\{v_{t'0}=\frac{m}{2},\, v_{t'j}=1\right\}_{j \in \mathcal{M}}\right), \quad t' = q+1.
\]
The customer chooses among the offered resources and the outside option according to the BAM model.

\paragraph{Value of \(\overline{V}(\mathcal{I}_m)\).}
Suppose that \(s\) of the first \(q\) requests are accepted. Their total revenue is \(s/(2m)\), and \(m-s\) resources remain available for the final request. Since all remaining resources have the same positive reward and attractiveness, it is optimal to offer all of them. Hence, conditional on \(s\), the total expected revenue is
\[
g(s) = \frac{s}{2m} + \frac{m-s}{m/2+m-s} = \frac{s}{2m} + \frac{m-s}{3m/2-s}.
\]
For every \(0 \le s \le q\),
\[
g(s) = \frac{3}{4} - \frac{(m-2s)^2}{4m(3m-2s)} \le \frac{3}{4}.
\]
Thus, no feasible policy, including a randomized policy, can obtain expected revenue greater than \(3/4\). Because \(m/2\) is an integer and \(m/2 \le q\), equality is attained by accepting exactly \(m/2\) of the first \(q\) requests. Therefore,
\[
\overline{V}(\mathcal{I}_m) = \frac{3}{4}.
\]

\paragraph{Value of \(\emph{\texttt{SBLP}}(\mathcal{I}_m)\).}
We construct a feasible solution to \eqref{eq:sblp}. For every \(1 \le t \le q\) and \(j \in \mathcal{M}\), set
\[
x_{tj}([1,1]) = 1-\frac{t-1}{m}, \qquad y_{tj}([1,1]) = \frac{1}{m}, \qquad y^0_{tj}([1,1]) = y^0_t = 0.
\]
For the final request \(t'=q+1\), set, for every \(j \in \mathcal{M}\),
\[
x_{t'j}([1,1]) = 1-\frac{q}{m} = \frac{m+2}{3m}, \qquad y_{t'j}([1,1]) = \frac{2}{3m}, \qquad y^0_{t'j}([1,1]) = y^0_{t'} = \frac{1}{3}.
\]
The \eqref{eq:boundary_assortment} and \eqref{eq:balance} constraints hold by construction. For each early request, \(y_{tj}([1,1]) \le x_{tj}([1,1])\), \(\sum_{j \in \mathcal{M}}y_{tj}([1,1])=1=p_t\), and the \eqref{eq:scale} and \eqref{eq:no_purchase} constraints hold because \(v_{t0}=y^0_{tj}([1,1])=y^0_t=0\). For the final request,
\[
y_{t'j}([1,1])+y^0_{t'j}([1,1]) = \frac{2}{3m}+\frac{1}{3} = \frac{m+2}{3m} = x_{t'j}([1,1]),
\]
\[
v_{t'0}y_{t'j}([1,1]) = \frac{m}{2}\cdot\frac{2}{3m} = \frac{1}{3} = v_{t'j}y^0_{t'j}([1,1]),
\]
and
\[
y^0_{t'}+\sum_{j \in \mathcal{M}}y_{t'j}([1,1]) = \frac{1}{3}+m\cdot\frac{2}{3m} = 1=p_{t'}.
\]
The \eqref{eq:no_purchase} constraints also hold at equality. Thus, the solution is feasible and has objective value
\[
\texttt{SBLP}(\mathcal{I}_m) \ge q\left(m\cdot\frac{1}{2m}\cdot\frac{1}{m}\right)+m\cdot\frac{2}{3m} = \frac{m-1}{3m}+\frac{2}{3} = 1-\frac{1}{3m}.
\]

\paragraph{Conclusion.}
Consequently,
\[
\frac{\overline{V}(\mathcal{I}_m)}{\texttt{SBLP}(\mathcal{I}_m)} \le \frac{3/4}{1-1/(3m)} = \frac{9m}{4(3m-1)}.
\]
Taking \(m \to \infty\) along integers satisfying \(m \equiv 4 \pmod 6\) gives
\[
\inf_{\mathcal{I}} \left\{ \frac{\overline{V}(\mathcal{I})}{\texttt{SBLP}(\mathcal{I})} \right\} \le \frac{3}{4}.
\]
\hfill \Halmos

\section{Omitted Details in Section~\ref{sec:general_arrivals}}

\subsection{Full Details of \eqref{eq:sblp_general}}
\label{appendix:full_lp_general}
Below we provide the full details of~\eqref{eq:sblp_general}. Here, $y^{(k)0}_t$ denotes the probability that customer $t$ of type $k$ arrives and does not choose any resource.
\begin{align}
\label{eq:sblp_general_appendix}
\tag{SBLP-General}
     \max_{\mathbf{x} \ge \mathbf{0}, \mathbf{y} \ge \mathbf{0}} \quad& \sum_{j \in \mathcal{M}}  \sum_{t \in \mathcal{T}} \sum_{1 \le a \le b \le N} \sum_{k=1}^{K_t} w^{(k)}_{tj} y^{(k)}_{tj}([a,b]) \\
    \label{eq:allocate_general_appendix}
    \tag{Online}
    \text{ s.t.} \qquad& y^{(k)0}_{tj}([a,b]) +  y^{(k)}_{tj}([a, b]) \le x_{tj}([a,b]) \cdot p^{(k)}_t, &&  \forall j \in \mathcal{M},  t \in \mathcal{T},  1 \le a \le b \le N, k \in [K_t] \\
    \label{eq:no_allocate_general}
    \tag{Feasibility}
    \quad&  y^{(k)0}_{tj}([a,b]) +  y^{(k)}_{tj}([a, b]) \le \mathbbm{1}\{ [l^{(k)}_t,r^{(k)}_t] \subseteq [a,b], v^{(k)}_{tj} > 0 \}, && \forall j \in \mathcal{M},  t \in \mathcal{T},  1 \le a \le b \le N, k \in [K_t] \\
    \quad& x_{tj}([a,b]) = x_{t-1,j}([a,b])  -  \sum_{k=1}^{K_{t-1}} y^{(k)}_{t-1,j}([a,b])   +
    \sum_{k=1}^{K_{t-1}} \sum_{ r^{(k)}_{t-1} \le b' \le N} &&  \mathbbm{1}\{ b+1 = l^{(k)}_{t-1} \}y^{(k)}_{t-1,j}([a, b']) \notag
    \\
    \label{eq:trans_general}
    \tag{Balance}
    & \quad + \sum_{k=1}^{K_{t-1}} \sum_{1 \le a' \le l^{(k)}_{t-1} } \mathbbm{1}\{ r^{(k)}_{t-1} = a-1 \} y^{(k)}_{t-1,j}([a',b]), && \forall j \in \mathcal{M},  2 \le t \le T,  1 \le a \le b \le N  \\
    \label{eq:boundary_general}
    \tag{Boundary}
    \quad& x_{1j}([a,b]) = \mathbbm{1}\{ [a,b]=[1,N]\}, &&  \forall j \in \mathcal{M},  1 \le a \le b \le N \\
    \label{eq:capacity_general}
    \tag{Capacity}
    \quad& y^{(k)0}_t + \sum_{j \in \mathcal{M}}\sum_{1 \le a \le b \le N} y^{(k)}_{tj}([a,b]) = p^{(k)}_t, &&   \forall t \in \mathcal{T}, k \in [K_t]. \\
    &
    \tag{Scale}
    \label{eq:scale_general}
     v^{(k)}_{t0}y^{(k)}_{tj}([a,b]) = v^{(k)}_{tj}y^{(k)0}_{tj}([a,b]), && \forall j \in \mathcal{M}, t  \in \mathcal{T}, 1 \le a \le b \le N, k \in [K_t]\\
    \tag{Opt-out}
    \label{eq:no_purchase_general}
    \quad&  \sum_{1 \le a \le b \le N} y^{(k)0}_{tj}([a,b]) \le y^{(k)0}_t, && \forall j \in \mathcal{M}, t \in \mathcal{T}, k \in [K_t].
\end{align}

\subsection{Details of \textsc{General-Random}}
\label{subsec:general_random}

In this section, we describe and analyze the coupling procedure \textsc{General-Random}, which can be regarded as a two-dimensional generalization of $\textsc{Random}$. Our explanation below interprets our idea of random variable coupling and then describes the implementation through Algorithm~\ref{alg:general_random}. Finally, we prove Proposition~\ref{prop:general_random}.

Suppose inputs $t, \tilde{k}, \tilde{j}$, $\boldsymbol{\phi}$, $\mathbf{Q}$ and $\mathbf{Q}'$ satisfy the conditions in Proposition~\ref{prop:general_random}.
For ease of notation, let $\theta_k$ denote the position of $k$ in the permutation $\boldsymbol{\phi}$, that is, $\phi_{\theta_k} = k$.
We aim to ensure that the outputs $\{k_j\}_{j \in \mathcal{M}}$ are independent across resources and that the marginal distribution of each $k_j$ satisfies $\Pr[k_j = k] = q_{kj}$.
Our coupling between $(\tilde{k},\tilde{j})$ and $\mathbf{k}$ builds on a standard technique for coupling two discrete random variables with a sequence of independent Bernoulli random variables. 
Define independent Bernoulli random variables $ \{ X_{kj} \}_{k \in [K_t], j \in \mathcal{M}}$, with 
\begin{align}
\label{eq:prob_X}
\Pr[X_{kj} = 1] = \frac{q'_{k,j}}{ 1 - \sum_{k': \theta_{k'} < \theta_k}\sum_{j' \in \mathcal{M}} q'_{k'j'} - \sum_{j' < j} q'_{k j'}}.
\end{align}

Then, we couple $(\tilde{k},\tilde{j})$ with $\{X_{kj}\}$ as follows:
\begin{align}
\label{eq:couple_tilde}
    (\tilde{k}, \tilde{j}) = (k, j)
    \quad\text{iff}\quad 
    X_{kj} = 1
    \ \text{and}\ 
    X_{k',j'} = 0
    \quad \forall (\theta_{k'}, j') \prec (\theta_k, j),
\end{align}
where $\prec$ denotes the lexicographic order.
It is easy to verify that such a coupling preserves the joint probabilities of $(\tilde{k}, \tilde{j})$ specified in \eqref{eq:distribution_general}.

For each $j \in \mathcal{M}$, we couple the variable $k_j$ with the
collection of Bernoulli indicators $\{X_{kj}\}_{k \in [K_t]}$ and an
auxiliary discrete random variable $Y_j$.  
The variable $Y_j$ is chosen to be independent of all other $Y$'s and all
$X$'s, and is defined by
\begin{align}
\label{eq:prob_Y}
    \Pr[Y_j = k]
    =
    \frac{\alpha_{kj}}{\sum_{k' \in [K_t]^+} \alpha_{k'j}},
    \qquad \forall k \in [K_t]^+,
\end{align}
where
\[
    \alpha_{kj}
    =
    q_{kj}
    \;-\;
    \Pr[X_{kj} = 1]\,
    \prod_{\;k' :\, \theta_{k'} < \theta_{k}}
        \Pr[X_{k'j} = 0],
    \qquad \forall k \in [K_t],
    \qquad\text{ and }\qquad
    \alpha_{0j} = q_{0j}.
\]
These probabilities are well defined because $\alpha_{kj} \ge 0$, which follows
from the fact that $\Pr[X_{kj}=1] \le q_{kj}$ (by
\eqref{eq:regular_condtion_general}).
We then define $k_j$ by the following coupling rule:
\begin{equation}
\label{eq:couple_j}
    k_j =
    \begin{cases}
        {\arg\min}_{k:X_{kj} = 1} \{ \theta_k\}, 
        & \text{if } \{ k \in [K_t] | X_{kj} = 1 \} \neq \emptyset, \\[2mm]
        Y_j, & \text{otherwise.}
    \end{cases}
\end{equation}
In words, $k_j$ is set to the \emph{earliest} request type (in the order
$\boldsymbol{\phi}$) for which $X_{kj}=1$, and if all $X_{kj}$ are zero, then $k_j$ is drawn from the fallback distribution $Y_j$.
It is straightforward to verify that each $k_j$ produced by
\eqref{eq:couple_j} has the correct marginal distribution, and that the
variables $\{k_j\}_{j \in \mathcal{M}}$ are mutually independent.
Consequently, the vector $\mathbf{k}$ satisfies the joint distribution
specified in~\eqref{eq:indendent_general}.
As such, the
random pair $(\tilde{k},\tilde{j})$ and the vector $\mathbf{k}$ are coupled
through the shared randomness of the $X$- and $Y$-variables.

Algorithm~\ref{alg:general_random} implements this coupling explicitly.  
On Lines~2--8, for any realized input $(\tilde{k}, \tilde{j})$ with 
$\tilde{k} \in [K_t]$ and $\tilde{j} \in \mathcal{M}$, the values of the 
variables $X_{kj}$ are sampled according to the conditional probabilities 
prescribed by~\eqref{eq:couple_tilde}:
\begin{align*}
    &\Pr[X_{kj} = 0 \mid \tilde{k}, \tilde{j}] = 1,
    &&\forall\, (\theta_k, j) \prec (\theta_{\tilde{k}}, \tilde{j}), \\[1mm]
    &\Pr[X_{\tilde{k}\tilde{j}} = 1 \mid \tilde{k}, \tilde{j}] = 1, \\[1mm]
    &\Pr[X_{kj} = x \mid \tilde{k}, \tilde{j}]
      = \Pr[X_{kj} = x],
    &&\forall\, x \in \{0,1\},\ \forall\, (\theta_k, j) \succ (\theta_{\tilde{k}}, \tilde{j}).
\end{align*}
Since the algorithm ultimately needs only the vector $\mathbf{k}$, whenever 
$X_{kj}$ is realized as the first value equal to~1 for resource~$j$, we set 
$k_j = k$ as specified in~\eqref{eq:couple_j}.  
There is one corner case: when either $\tilde{k} \notin [K_t]$ or 
$\tilde{j} \notin \mathcal{M}$.  
Because we do not require the equality 
$\sum_{k \in [K_t],\, j \in \mathcal{M}} q'_{kj} = 1$, this situation may occur with nonzero probability.  
In this case, we simply set all $X_{kj} = 0$ and the algorithm proceeds directly to Line~9.
On Lines~9--10, for each resource $j$ such that 
$\{ k \in [K_t] : X_{kj} = 1 \} = \emptyset$ (equivalently, $k_j = 0$ at that point), 
we sample $Y_j$ and assign $k_j = Y_j$, again following the rule in~\eqref{eq:couple_j}.

\begin{algorithm}
\caption{\textsc{General-Random}$(t, \tilde{k}, \tilde{j}, \boldsymbol{\phi}, \mathbf{Q}, \mathbf{Q}')$ }
\label{alg:general_random}
\begin{algorithmic}[1]
    \REQUIRE Period $t$, a request type $\tilde{k}$, a customer choice $\tilde{j}$, a permutation $\boldsymbol{\phi}$, two matrices $\mathbf{Q}$ and $\mathbf{Q}'$.
    \ENSURE $\mathbf{k}=[k_j]_{j \in \mathcal{M}}$.
     \STATE Initialize $k_j \gets 0$ for all $j \in \mathcal{M}$.
    \IF{$\tilde{k} \in [K_t]$ and $\tilde{j} \in \mathcal{M}$}
    \FOR{$i = K_t$ downto $1$}
    \FOR{$ j = 1$ to $M$}
    \IF{$(i, j )= (\theta_{\tilde{k}}, \tilde{j}) $}
    \STATE $k_{\tilde{j}} = \tilde{k}$.
    \ELSIF{$(i, j )\succ
    (\theta_{\tilde{k}}, \tilde{j}) $}
    \STATE With probability $\Pr[X_{\phi_i, j} = 1]$ defined in \eqref{eq:prob_X}, $k_j \leftarrow \phi_i$.
    \ENDIF
    \ENDFOR
    \ENDFOR
    \ENDIF

    \FOR{each $j \in \mathcal{M}$ with $k_j = 0$}
        \STATE $k_j\leftarrow Y_j$, where $Y_j$ is sampled according to probabilities defined in \eqref{eq:prob_Y}.
    \ENDFOR
\end{algorithmic}
\end{algorithm}

Next, we proceed to prove Proposition~\ref{prop:general_random}.

\proof{Proof of Proposition~\ref{prop:general_random}.}

For the complexity bound, the nested loops on Lines~3--4 run in 
$O(MK_t)$ time, and the loop on Line~9 iterates over all $j \in \mathcal{M}$ 
and requires $O(K_t)$ time for each sample.  
Thus, the overall running time of the procedure is $O(MK_t)$.

The feasibility property follows immediately from Line~6.

The independence property follows directly from the coupling construction:
each $k_j$ is generated solely from the random variables 
$\{X_{kj}\}_{k \in [K_t]}$ and an independent auxiliary variable $Y_j$, and all
$X$'s and $Y$'s are mutually independent across different resources.  
Therefore, the components of $\mathbf{k}$ are independent, and each has the 
correct marginal distribution.
\hfill \Halmos
 


\subsection{Omitted Proofs}
\label{subsec:general_arrival_proof}

\proof{Proof of Proposition~\ref{prop:virtual_proerty_general}.}
The proof proceeds by the same induction used for 
Proposition~\ref{prop:virtual_property_assortment}, with the necessary 
adaptations to handle general arrivals.  
Suppose in some period~$t'$ the induction hypothesis holds:  
Property~\ref{property:virtual} is satisfied for all virtual statuses 
$\{ \bar{\mathbf{s}}^j_t \}_{t \le t'}$, and all offered assortments up to period~$t'$ are feasible.

In period $t'$, 
\eqref{eq:regular_condtion_general} is satisfied as shown in \eqref{eq:satisfy_general_random_condition}.
In addition, by Line~12 of Algorithm~\ref{alg:assortment_general},
$({k}^*,{j}^*)$ follows the distribution in \eqref{eq:distribution_general} if $\mathbf{Q}'$ is defined as in \eqref{eq:dfn_Q'}.
So, we are able to utilize Proposition~\ref{prop:general_random} to complete the proof.
Whenever resource $j^*$ allocates interval $[l^{(k^*)}_{t'}, r^{(k^*)}_{t'}]$ to the 
realized request~$t'$ (Line~12), the algorithm correspondingly updates the virtual status via
\(
    \bar{\mathbf{s}}^{j}_{t'+1}
    \gets 
    \bar{\mathbf{s}}^j_{t'} - \mathbf{1}_{[l^{(k^*)}_{t'},\, r^{(k^*)}_{t'}]}
\)
for $j = j^*$, which is from the feasibility property in Proposition~\ref{prop:general_random}.  
Thus, the Lower Bound property $\bar{\mathbf{s}}^{j}_{t'+1} \le \mathbf{s}^j_{t'+1}$ holds.
The feasibility of all offered assortments in period $t'+1$ follows directly from 
the Lower Bound property.
Finally, the Marginal and Independence conditions for the virtual status in period $t'+1$ follow immediately from the independence property in 
Proposition~\ref{prop:general_random}.
This completes the induction and the proof.
\hfill \Halmos

\proof{Proof of Theorem~\ref{thm:BAM_general}.}
We can also use the analysis \eqref{eq:lower_bound_S} for the expected revenue of offering assortment $\mathcal{S}$ to customer $t$ in the proof of Theorem~\ref{thm:1/8} to get that for the assortment $\mathcal{S}^{(k)}$. In particular, for any period $t \in \mathcal{T}$ and type $k \in [K_t]$, we have 
\begin{align}
    \minor{\mathbb{E}[R_t^{(k)}] \ge 0.271 \sum_{j \in \mathcal{M}} \frac{w^{(k)}_{tj} \sum_{a,b}y^{(k)}_{tj}([a,b])}{p^{(k)}_t}.}
\end{align}
Combining with Lemma~\ref{lemma:general 1 - 1/e}, we immediately get that the expected revenue from period $t$ is at least \minor{$0.271(1-1/e)\sum_{k \in [K_t], j \in \mathcal{M}} w^{(k)}_{tj} \sum_{a,b}y^{(k)}_{tj}([a,b])$}. Hence, \minor{using} Lemma~\ref{lemma:upper_general}, we get that \minor{$V^\pi(\mathcal{I}) \ge 0.271(1-1/e) \overline{V}(\mathcal{I})$} if $\pi$ denotes Algorithm~\ref{alg:assortment_general}.
\hfill \Halmos

\proof{Proof of Theorem~\ref{thm:accept_or_reject_general}.}
We connect Algorithm~\ref{alg:assortment_general} to Algorithm~\ref{alg:single_item} so that we can borrow the analysis. Note that $\mathcal{S}^{(k)}$ defined in \eqref{eq:s_from_p_general} is indeed selecting the resource with highest reward if $v^{(k)}_{t0} = 0$ and hence in Line 12 of Algorithm~\ref{alg:assortment_general}, we actually allocate the resource with highest reward from $\mathcal{P}^{(k)}$ to the type-$k$ customer $t$. This is actually the same as Lines 6-8 in Algorithm~\ref{alg:single_item}. Hence, we can apply the analysis \eqref{eq:lower_bound_revenue_period} for $\mathbb{E}[R_t]$ to $\mathbb{E}[R_{t}^{(k)}]$ to derive that 
\[
\mathbb{E}[R_{t}^{(k)}] \ge (1-\frac{1}{e})  \sum_{j \in \mathcal{M}} \frac{w^{(k)}_{tj} \sum_{a,b}y^{(k)}_{tj}([a,b])}{p^{(k)}_t}.
\]
Hence, combining with Lemma~\ref{lemma:general 1 - 1/e}, we get that the expected revenue from period $t$ is at least $(1-1/e)^2\sum_{k \in [K_t], j \in \mathcal{M}} w^{(k)}_{tj} \sum_{a,b}y^{(k)}_{tj}([a,b])$. Using Lemma~\ref{lemma:upper_general}, we get that $V^\pi(\mathcal{I}) \ge (1-1/e)^2\overline{V}(\mathcal{I})$ if $\pi$ is the Algorithm~\ref{alg:assortment_general} and $\mathcal{I}$ is an instance of BAM-based scenario with $v^{(k)}_{t0} = 0$.
\hfill \Halmos

\end{document}